\documentclass[aps,prc,preprint,groupedaddress,showpacs,superscriptaddress,floatfix]{revtex4-1}

\usepackage{graphicx,amsmath,amssymb,booktabs,dcolumn}
\usepackage{braket}
\usepackage[colorlinks,linkcolor=blue,urlcolor=blue,citecolor=green]{hyperref}
\usepackage[caption=false,farskip=5pt,captionskip=1pt]{subfig}

\newcommand{\diff}{\ensuremath\mathrm{d}}
\renewcommand{\vec}[1]{\ensuremath\boldsymbol{#1}}
\newcommand{\GeV}{\ensuremath\mathrm{GeV}}
\newcommand{\udarrow}{\ensuremath{{\uparrow,\downarrow}}}
\newcommand{\duarrow}{\ensuremath{{\downarrow,\uparrow}}}
\newcommand{\uuarrow}{\ensuremath{\uparrow,\uparrow}}

\begin{document}
\title{Electroweak properties of octet baryons in light-cone quark-diquark model}

\newcommand*{\PKU}{School of Physics and State Key Laboratory of Nuclear Physics and
Technology, Peking University, Beijing 100871,
China}\affiliation{\PKU}
\newcommand*{\CICQM}{Collaborative Innovation
Center of Quantum Matter, Beijing, China}\affiliation{\CICQM}
\newcommand*{\CHEP}{Center for High Energy
Physics, Peking University, Beijing 100871,
China}\affiliation{\CHEP}

\author{Jun~Zhang}\affiliation{\PKU}
\author{Bo-Qiang~Ma}\email{mabq@pku.edu.cn}\affiliation{\PKU}\affiliation{\CICQM}\affiliation{\CHEP}

\begin{abstract}
We study the electroweak properties of ground state octet baryons in a relativistic quark-spectator-diquark model, with light-front formalism applied to take relativistic effects into account.
Our model provides a consistent picture of the electroweak properties of the ground state octet baryons in the low momentum transfer region.
The Melosh-Wigner rotation is applied as the transformation relation between spinors in the instant form and front form.
Numerical results are presented for the magnetic moments, weak transition charges and Sachs form factors.
Our results are in good agreement with experimental measurements and other theoretical results.
\end{abstract}

\maketitle

\section{Introduction}
The well studied Quantum Chromodynamics (QCD) is the theory of the strong interaction which binds quarks through gluons to form hadrons.
However, the fundamental picture of hadron structure in terms of QCD is still a mystery because of its non-perturbative nature.
The coupling constants grow large in low-$Q^2$ domain, rendering the traditional perturbative method powerless.
This non-perturbative behavior makes the calculation of hadron properties, such as masses, form factors, an extremely difficult problem pending a solution.
Over the last few decades, lattice gauge theory~\cite{wilson_confinement_1974} has undergone steady development.
Progress has been achieved in the calculation of baryon masses~\cite{durr_ab-initio_2008,*walker-loud_light_2009,*young_octet_2010} as well as electroweak properties~\cite{lin_first_2009,shanahan_electric_2014,*shanahan_magnetic_2014}.
Another important non-perturbative method utilizes the Dyson-Schwinger equations (DSE)~\cite{dyson_$s$_1949,*schwinger_greens_1951}.
Investigations into baryon masses, form factors, PDFs and GPDs have been carried out in this approach~\cite{oettel_octet_1998,*eichmann_nucleon_2010} and provide us valuable insights.
No less important are the phenomenological models which capture the underlying dynamical features of the baryons.
They are able to produce good results and predictions with only a few assumptions.
In this work we are to examine the electroweak form factors of the octet baryons with a light-front quark-diquark model.

The electromagnetic form factors provide information on the internal momentum space distribution of the electric and magnetic charge, thus granting us a chance to glimpse into the substructure of baryons.
In the 2000s, polarization transfer experiments~\cite{the_jefferson_lab_hall_a_collaboration_ratio_2000,*gayou_measurements_2001,*puckett_recoil_2010} revealed that the ratio between the two Sachs from factors of proton, namely $\mu_p G_E^p(Q^2)/G_M^p(Q^2)$, decreases almost linearly with $Q^2$, in contrast to what we assumed to be constant.
It was shown that our model is able to reproduce such behavior~\cite{ma_electromagnetic_2002}.
Recent muonic hydrogen experiments~\cite{pohl_size_2010,*antognini_proton_2013} find that the proton charge radius is measured to be $4\%$ smaller than the results from electron-proton scattering and Lamb shift in electronic hydrogen experiments.
The origin of this deviation is still unresolved.

The diquark describes a two-quark system.
This concept helps us to understand nucleon properties and high-energy collisions.
In the DSE approach, diquark leads to important simplifications which are used extensively in the literature.
This idea was first mentioned by Gell-Mann~\cite{gell-mann_schematic_1964} and then developed by Ida~\cite{ida_baryon_1966} and Lichtenberg~\cite{lichtenberg_baryon_1967}.
The light-front quark-diquark model further extends this picture in the light-front formalism to take relativistic effects into account.
The light-front (LF) formalism, or the ``Front Form''~\cite{dirac_forms_1949}, offers a natural framework to describe hadron structure in terms of their quark and gluon degrees of freedom.
The fields are quantized at fixed LF time $x^+=x^0+x^3$ instead of the standard time $x^0$.
The LF formalism maximizes the number of kinematic generators of the Poincar\'e group, making it ideal for dealing with relativistic systems.
The simple structure of the LF vacuum enables the hadron to be described as an unambiguous frame-independent n-particle Fock-state expansion, with the coefficients called light front wave functions (LFWF)~\cite{brodsky_quantum_1998}.
This formalism has been employed to solve the proton spin puzzle~\cite{ma_melosh_1991,*ma_proton_1993}.
The recently developed light-front holographic method which combines LF quantization with Ads/CFT duality successfully reproduces the Regge trajectories of hadronic spectrum~\cite{karch_linear_2006,brodsky_ads/cft_2008}.

In this work we investigate the electromagnetic and weak transition form factors of the octet baryons in this light-front quark-spectator-diquark model.
The numerical results of electromagnetic form factors, magnetic moments and semi-leptonic decay parameters are presented and compared with experimental results.
Sec.~\ref{sec:model} is a brief review of the light-front quark-diquark model.
We calculate the form factors and magnetic moments in Sec.~\ref{sec:form_factor} and axial charges for the semi-leptonic beta decays in Sec.~\ref{sec:beta_decay} within this framework.
The numerical results are presented in Sec.~\ref{sec:numerical}.
Our investigation is summarized in Sec.~\ref{sec:summary}.

\section{Light-front quark-spectator-diquark model for hadrons\label{sec:model}}
In the light-front formalism, the hadron state can be expressed as an expansion with n-particle Fock states as the basis~\cite{brodsky_quantum_1998}:
\begin{multline}
  \Ket{H: P^+, \vec{P}_\perp, \lambda} = \sum_{n,\{\lambda_i\}} \prod^N_{i=1} \int
  \frac{\diff x_i \diff^2 \vec{k}_{\perp i}}{2{(2\pi)}^3 } (16\pi^3)
  \delta \left(1-\sum_{j=1}^N x_j \right) \delta^{(2)} \left( \sum_{j=1}^N \vec{k}_{\perp j} \right) \\
  \times \psi_{n/H} ( \{x_i\}, \{\vec{k}_{\perp j} \}, \{\lambda_j\} ) \Ket{n: \{x_i P^+\}, \{x_i \vec{P}_\perp + \vec{k}_{\perp i}\}, \{\lambda_i\}},
\label{eq:expansion}
\end{multline}
where $\ket{n}$ are the light-front Fock states quantized at fixed light-front time $x^+ = x^0+x^3$.
$x_i=k_i^+/P^+$, $\vec{k}_{\perp i}$ and $\lambda_i$ are the light-cone momentum fraction, intrinsic transverse momentum and helicity of the $i$th component respectively.
The coefficients $\psi_{n/H}(\{x_i\}, \{\vec{k}_{\perp i}\}, \{\lambda_i\})$ are the LFWFs.
In principle, they can be obtained by solving the light-cone Hamiltonian equation $(H_{LC}-M^2)\ket{B}=0$ although the solution in 4 dimensions is still not achievable at the moment.

Fortunately, the $\mathrm{SU}(6)$ quark-diquark model provides a way to approximate these wave functions.
In the impulse approximation, a single constituent quark in the baryon is struck by the incident lepton, while the remaining part is treated as a quasi-particle spectator which provides other quantum numbers of the baryon.
Thus it is natural and convenient to treat the baryon as a quark-spectator-diquark system.
In our model, the baryon is described as a constituent quark involved in the interaction and a spectator diquark which serves to provide the quantum numbers.
It is also possible to absorb some non-perturbative effects into the diquark parameters.
This model has been proved to be a reliable tool in the calculation of quark helicity and transversity distributions~\cite{ma_x-dependent_1996,ma_quark_1998}, form factors~\cite{ma_electromagnetic_2002,ma_axial_2002,liu_generalized_2014}, transverse momentum dependent parton distributions~\cite{lu_sivers_2004}, Wigner distributions~\cite{liu_quark_2015}, etc.
Nevertheless, we do need to point out that this picture is actually too simplistic.
More realistic analysis is necessary to achieve a better description of baryon structures.

For a baryon, the instant form $\mathrm{SU}(6)$ quark-diquark wave function is written as
\begin{equation}
  \Ket{B}^{\udarrow} = 
  \cos\theta \sum_q a_q \Ket{q_{1}S(q_{2}q_{3})}_T^{\udarrow}
  + \sin\theta \sum_{q'} b_{q'} \Ket{q'_{1}V(q'_{2}q'_{3})}_T^{\udarrow},
  \label{eq:quark_diquark}
\end{equation}
in which $\sum_q$ and $\sum_{q'}$ stand for the sum of different quark-diquark components, $\ket{q_1 S(q_2 q_3)}_T$ represents a Fock state with a quark $q_1$ and a scalar diquark $S(q_2, q_3)$, respectively $\ket{q'_1 V(q'_2 q'_3)}_T$ represents a Fock state with a quark $q'_1$ and an axial-vector diquark $V(q'_2, q'_3)$.
The subscript $T$ means these are the instant form Fock states.
$a_q$ and $b_{q'}$ are coefficients from the $\mathrm{SU}(6)$ symmetry satisfying normalization condition $\cos^2\theta \sum_q a_q^2 + \sin^2\theta \sum_{q'} b_{q'}^2 = 1$. 
The $a_q$ and $b_{q'}$ for all the octet baryons can be found in Ref.~\cite{lichtenberg_quark-diquark_1968}.
$\theta$ is the mixing angle related to spin-flavor $\mathrm{SU}(6)$ symmetry breaking.
The symmetric case $\theta=\pi/4$ is chosen as in our previous works~\cite{ma_x-dependent_1996,ma_quark_1998,ma_electromagnetic_2002,ma_axial_2002,liu_generalized_2014,lu_sivers_2004,liu_quark_2015}.
We consider only the Fock states with zero orbital angular momentum, thus the state of a quark-diquark system with a certain helicity can be expanded to a quark state together with a diquark state as
\begin{align}
  \ket{q_1 S(q_2 q_3)}_T^\udarrow =&\, 
  \int\frac{\diff x \diff^2 \vec{k}_{\perp}}{16\pi^3 }\,
  \varphi_{q_1 S(q_2 q_3)}(x, \vec{k}_\perp) \ket{q_1^\udarrow S(q_2 q_3)}_T,
  \label{eq:fock_expansion_1} \\
  \ket{q_1 V(q_2 q_3)}_T^\udarrow =&
  \pm \int\frac{\diff x \diff^2 \vec{k}_{\perp}}{16\pi^3}
  \frac{1}{\sqrt{3}} \varphi_{q_1 V(q_2 q_3)}(x, \vec{k}_\perp) \nonumber\\
  &\,\big[ - \ket{q_1^\udarrow V^0(q_2 q_3)}_T 
  + \sqrt{2} \ket{q_1^\duarrow V^{\pm 1}(q_2 q_3)}_T \big],
  \label{eq:fock_expansion_2}
\end{align}
where $\varphi_{q_1 S(q_2 q_3)}(x, \vec{k}_\perp)$ and $\varphi_{q_1 V(q_2 q_3)}(x, \vec{k}_\perp)$ are the momentum space wave functions corresponding to each Fock state, in which $x$ is the light-cone momentum fraction and $\vec{k}_\perp$ is the intrinsic transverse momentum of the quark.
Meanwhile the spin-flavor structures are encoded in $\ket{q_1^\udarrow S(q_2 q_3)}_T$, $\ket{q_1^\udarrow V^0(q_2 q_3)}_T$ and $\ket{q_1^\duarrow V^{\pm 1}(q_2 q_3)}_T$.
For the example of a proton, the instant form wave function in Eq.~\eqref{eq:quark_diquark} can be written as
\begin{equation}
  \ket{p}^\udarrow = \frac{1}{\sqrt{2}} \ket{uS(ud)}_T^\udarrow
  - \frac{1}{\sqrt{6}} \ket{uV(ud)}_T^\udarrow
  + \frac{1}{\sqrt{3}} \ket{dV(uu)}_T^\udarrow,
  \label{eq:proton_spin_wf}
\end{equation}
with the Fock states expanded with Eqs.~\eqref{eq:fock_expansion_1} and~\eqref{eq:fock_expansion_2}.
Other octet baryons differ from proton only by the coefficients $a_i$, $b_i$ and quark components in Eq.~\eqref{eq:quark_diquark}. 
Their wave functions can be obtained in a similar fashion by referring to Ref.~\cite{lichtenberg_quark-diquark_1968}.

With the instant form wave functions available, the LFWFs can be obtained by transforming the instant form spinors to the front form with the Melosh-Wigner rotation~\cite{wigner_unitary_1939,*melosh_quarks:_1974,*[{See also }] buccella_current_1974}.
For the quark spinors in Eqs.~\eqref{eq:fock_expansion_1}-\eqref{eq:fock_expansion_2} we have
\begin{equation}
  \begin{bmatrix}
    q_T^\uparrow \\ 
    q_T^\downarrow
  \end{bmatrix}
  = W_q
  \begin{bmatrix}
    q_F^\uparrow \\
    q_F^\downarrow
  \end{bmatrix},
  \label{eq:wigner_rotation_spinor}
\end{equation}
where $q_T$ and $q_F$ denote the instant form and front form spinors respectively.
The transformation matrix is~\cite{melosh_quarks:_1974}
\begin{equation}
  W_q = \omega
  \begin{bmatrix}
    k^++m & -k^R \\
    k^L & k^++m
  \end{bmatrix},
\label{eq:W_q}
\end{equation}
where $m$ is quark mass, $\omega = 1/ \sqrt{(k^++m)^2+k^Lk^R}$ is the normalization factor, and $k^+=k^0+k^3$, $k^{R,L}=k^1\pm k^2$.
The scalar diquarks remain the same in both forms because of their zero spin.
The Wigner rotation for the axial-vector diquark spinors is
\begin{equation}
  \begin{bmatrix}
    V_T^1\\ V_T^0\\ V_T^{-1}
  \end{bmatrix}
  = W_V
  \begin{bmatrix}
    V_F^1\\ V_F^0\\ V_F^{-1}
  \end{bmatrix},
\label{eq:wigner_rotation_vector}
\end{equation}
in which the transformation matrix is provided as~\cite{ahluwalia_front_1993}
\begin{equation}
  W_V = \omega_V^2
  \begin{bmatrix}
    (k_V^++m_V)^2 & -\sqrt{2} (k_V^++m_V)k_V^R & (k_V^R)^2\\
    \sqrt{2} (k_V^++m_V)k_V^L & (k_V^++m_V)^2-k_V^L k_V^R & 
    -\sqrt{2} (k_V^++m_V)k_V^R\\
    (k_V^L)^2 & \sqrt{2}(k_V^++m_V)k^L & (k_V^++m_V)^2
  \end{bmatrix},
\label{eq:W_V}
\end{equation}
where $m_V$ is diquark mass and $\omega_V = 1/ \sqrt{(k_V^++m_V)^2+k_V^L k_V^R}$. $k_V^+$ and $k_V^{R,L}$ have similar definitions as in Eq.~\eqref{eq:wigner_rotation_spinor}.
It has been verified that spinors obtained by Wigner rotation are equivalent to those from direct calculation in the light front formalism~\cite{lepage_exclusive_1980}.

Following the above discussion, the light-cone wave function of octet baryons in Eq.~\eqref{eq:expansion} can be approximated in the $\mathrm{SU}(6)$ quark-diquark formalism by transforming the instant form spinors in Eqs.~\eqref{eq:fock_expansion_1} and \eqref{eq:fock_expansion_2} to front form spinors using Eq.~\eqref{eq:wigner_rotation_spinor} and Eq.~\eqref{eq:wigner_rotation_vector}.
We have after this transformation
\begin{align}
  \ket{qS}_T^\lambda &= \int\frac{\diff x \diff^2 \vec{k}_{\perp}}{16\pi^3 }\,
  \sum_{i=\{\udarrow\}}A^{i,\lambda}_{qS}\ket{q^i S}_F, \label{eq:felf_1} \\
  \ket{qV}_T^\lambda &= \int\frac{\diff x \diff^2 \vec{k}_{\perp}}{16\pi^3 }\,
  \sum_{i=\{\udarrow\}}\,\sum_{j=\{-1,0,1\}} B^{ij,\lambda}_{qV} \ket{q^i V^j}_F, \label{eq:felf_2}
\end{align}
in which $\lambda=\,\uparrow \mbox{or} \downarrow$, and
\begin{align}
  A^{i,\lambda}_{qS} &= W_{q}^{\lambda,i} \varphi_{qS}\\
  B^{ij,\uparrow}_{qV} &= \left( -\sqrt{\frac{1}{3}} W_{q}^{\uparrow,i} W_{V}^{0,j}
  + \sqrt{\frac{2}{3}} W_{q}^{\downarrow,i} W_{V}^{1,j} \right) \varphi_{qV}\\
  B^{ij,\downarrow}_{qV} &= \left( -\sqrt{\frac{1}{3}} W_{q}^{\downarrow,i} W_{V}^{0,j}
  + \sqrt{\frac{2}{3}} W_{q}^{\uparrow,i} W_{V}^{-1,j} \right) \varphi_{qV},
\end{align}
where $W_{q}^{i,j}$ and $W_{V}^{i,j}$ are elements of the matrices in Eq.~\eqref{eq:W_q} and Eq.~\eqref{eq:W_V}, $\varphi_{qS}$ and $\varphi_{qV}$ are the momentum space wave functions.
Compared with Eqs.~\eqref{eq:fock_expansion_1}-\eqref{eq:fock_expansion_2}, the components $q_2$, $q_3$ of diquark as well as the parameters of the matrix elements and momentum space wave functions are omitted for compactness.
By replacing the instant form Fock states in Eq.~\eqref{eq:quark_diquark} with the front form ones using Eqs.~\eqref{eq:felf_1}-\eqref{eq:felf_2}, we derive the Fock state expansion of octet baryons in front form, written in the form of Eq.~\eqref{eq:expansion} as:
\begin{align}
  \ket{B:P^+, \vec{P}_\perp = 0, \lambda} =&\,\int\frac{\diff^2 \vec{k}_\perp \diff x}{16\pi^3} 
  \Big[ \cos\theta \sum_q \sum_{i=\{\udarrow\}} a_q\,A^{i,\lambda}_{q_1S(q_2q_3)} \ket{q^i_1S(q_2q_3)} \nonumber\\
  &\,+ \sin\theta \sum_{q'} \sum_{i=\{\udarrow\}}\sum_{j=\{-1,0,1\}} b_{q'} B^{ij,\lambda}_{q'_1V(q'_2q'_3)} \ket{q'^i_1V^j(q'_2q'_3)} \Big].
\end{align}
The desired LFWFs can then be identified as the coefficients of each light-front Fock state:
\begin{align}
  \psi_{q_1S(q_2q_3)}^\lambda(x, \vec{k}_\perp, \lambda_{q_1}=i)  &= a_q\,\cos\theta\,A^{i, \lambda}_{q_1S(q_2q_3)}, \label{eq:lfwf1}\\
  \psi_{q_1V(q_2q_3)}^\lambda(x, \vec{k}_\perp, \lambda_{q_1}=i, \lambda_V=j) &= b_q\,\sin\theta\,B^{ij, \lambda}_{q_1V(q_2q_3)}.\label{eq:lfwf2}
\end{align}
LFWFs in our framework are composed of factors $A^{i,\lambda}_{qS}$ and $B^{ij,\lambda}_{qV}$ together with coefficients $a_q$ and $b_q$.
The factor themselves are expressed as matrix elements from $W_q$ and $W_V$ combined with the momentum space wave function $\varphi_{qD}(x,\vec{k}_\perp)$.
For the example of a proton, the light-front Fock state expansion is
\begin{align}
  \Ket{p:P^+,\vec{P}_\perp,\lambda} =&\,\int\frac{\diff^2 \vec{k}_\perp \diff x}{16\pi^3} 
  \,\Big[ \frac{1}{\sqrt{2}}\, \sum_{i=\{\uparrow,\downarrow\}}
  A^{i,\lambda}_{uS(ud)}\Ket{u^i S(ud)}_F \nonumber \\
  &\,- \frac{1}{\sqrt{6}}\,\sum_{i=\{\uparrow, \downarrow\}}\,\sum_{j=\{-1, 0, 1\}}
  B^{ij,\lambda}_{uV(ud)} \Ket{u^i V^j(ud)}_F \nonumber \\
  &\,+ \frac{1}{\sqrt{3}}\,\sum_{i=\{\uparrow, \downarrow\}}\,\sum_{j=\{-1, 0, 1\}}
  B^{ij,\lambda}_{dV(uu)} \Ket{d^i V^j(uu)}_F\Big]. \label{eq:plfexp}
\end{align}
The LFWFs $\psi_{n/H} (x, \vec{k}_{\perp}, \lambda_q)$ in Eq.~\eqref{eq:expansion} then can be identified as:
\begin{align}
  \psi_{uS(ud)}^\lambda(x, \vec{k}_\perp, \lambda_u=i)  &= \frac{1}{\sqrt{2}} A^{i, \lambda}_{uS(ud)}, \label{eq:plfwf_1}\\
  \psi_{uV(ud)}^\lambda(x, \vec{k}_\perp, \lambda_u=i, \lambda_V=j) &= -\frac{1}{\sqrt{6}} B^{ij, \lambda}_{uV(ud)},\\
  \psi_{dV(uu)}^\lambda(x, \vec{k}_\perp, \lambda_d=i, \lambda_V=j) &= \frac{1}{\sqrt{3}} B^{ij, \lambda}_{dV(uu)}. \label{eq:plfwf_3}
\end{align}
The expansions for other octet baryons have the same structure while the coefficients and quark-diquark components are different.

For the momentum space light-cone wave function, one choice is the Brodsky-Huang-Lepage (BHL) prescription~\cite{brodsky_hadronic_1981}:
\begin{equation}
  \varphi_{qD}(x, \vec{k}_\perp) = A_{qD} \exp\left(-\frac{\mathcal{M}^2}{8\beta_{D}^2}  \right),
    \label{eq:mswf_bhl}
\end{equation}
in which $D$ stands for the diquark with $S$ for the scalar and $V$ for the axial-vector.
$\beta_{qD}$ is a parameter signifying the transverse momentum scale and $A_{qD}$ is the normalization factor.
$\mathcal{M}$ is the invariant mass
\begin{equation}
  \mathcal{M} = \sqrt{\frac{m_q^2+\vec{k}^2_\perp}{x}+\frac{m_D^2+\vec{k}^2_\perp}{1-x}},
  \label{eq:invariant_mass}
\end{equation}
where $m_q$ and $m_D$ are the masses of the quark and spectator diquark respectively.
Other choices are also available, such as the power-law form wave function~\cite{brodsky_wave_1994}
\begin{equation}
  \varphi_{qD}(x, \vec{k}_\perp) = A_{qD} {\left(1 + \frac{\mathcal{M}^2}{8\beta_{D}^2}\right)}^{-3.5}.
  \label{eq:mswf_pl}
\end{equation}
In the low $Q^2$ region, this form and the BHL prescription usually give identical results.

\section{Electromagnetic form factors\label{sec:form_factor}}
The Dirac and Pauli form factors $F_1(Q^2)$ and $F_2(Q^2)$ are defined in terms of the electromagnetic current as:
\begin{equation}
  \Braket{P'\lambda'| J^{\mu} (0)| P\lambda} = \bar{u}_{\lambda'}(P') \left[ F_1(Q^2)\gamma^\mu
    + F_2(Q^2) \frac{i\sigma^{\mu\nu}q_\nu}{2M}\right]u_\lambda (P),
    \label{eq:ff_pauli_dirac}
\end{equation}
where $q^\mu={(P'-P)}^\mu$ is the momentum transfer, $Q^2=-q^2$, $\lambda$ and $\lambda'$ represent the initial and final baryon helicity respectively, the current $J^\mu = \bar{q}e\gamma^\mu q$.
Experimental results are usually extracted in the form of electric and magnetic Sachs form factors, which are defined as combinations of Dirac and Pauli form factors~\cite{sachs_high-energy_1962}:
\begin{align}
  G_E(Q^2) &= F_1(Q^2) - \frac{Q^2}{4M^2} F_2(Q^2), \label{eq:ff_sachs_1}\\
  G_M(Q^2) &= F_1(Q^2) + F_2(Q^2). \label{eq:ff_sachs_2}
\end{align}
The magnetic moment of baryon is simply the magnetic form factor in the $Q^2\to 0$ limit:
\begin{equation}
  \mu = G_M(0).
  \label{eq:ff_mu}
\end{equation}

The Dirac and Pauli form factors can be obtained by examining the plus component of the current:
\begin{align}
  F_1(Q^2) &= \Braket{P',\uparrow | \frac{J^+(0)}{2P^+} | P, \uparrow}, \label{eq:ff_plus_1}\\
  -q^L \frac{F_2(Q^2)}{2M} &= \Braket{P',\uparrow | \frac{J^+(0)}{2P^+} | P, \downarrow}. \label{eq:ff_plus_2}
\end{align}
We choose the frame in which
\begin{align}
  q &= (q^+, q^-, \vec{q}_\perp) = (0, \frac{-q^2}{P^+}, \vec{q}_\perp), \label{eq:dy_frame_1}\\
  P &= (P^+, P^-, \vec{P}_\perp) = (P^+, \frac{M^2}{P^+}, \vec{0}_\perp). \label{eq:dy_frame_2}
\end{align}
It is shown by Drell and Yan that the calculation of current matrix elements at space-like photon momentum can be simplified in such a frame~\cite{drell_connection_1970}. 
The expression of form factors in terms of LFWFs can be derived by sandwiching the electro-magnetic current with baryon states in Eq.~\eqref{eq:expansion} following Eqs.~\eqref{eq:ff_plus_1}-\eqref{eq:ff_plus_2}. After a little algebra, the form factors can then be expressed as overlaps of light-cone wave functions:
\begin{align}
  F_1(Q^2) &= \sum_a \int \frac{\diff^2 \vec{k}_\perp \diff x}{16\pi^3}
  e_q \psi_a^{\uparrow *}(x, \vec{k}'_{\perp}, \lambda)
  \psi_a^\uparrow (x, \vec{k}_{\perp}, \lambda), \label{eq:ffc_1}\\
  -q^L \frac{F_2(Q^2)}{2M} &= \sum_a \int \frac{\diff^2 \vec{k}_\perp \diff x}{16\pi^3}
  e_q \psi_a^{\uparrow *}(x, \vec{k}'_{\perp}, \lambda)
  \psi_a^\downarrow (x, \vec{k}_{\perp}, \lambda), \label{eq:ffc_2}
\end{align}
where $\sum_a$ is the sum over different quark-diquark components, $e_q$ is the charge of the struck constituent, the light-cone wave functions $\psi_a^{\uparrow}(x, \vec{k}_{\perp}, \lambda)$ are given by Eq.~\eqref{eq:lfwf1} and~\eqref{eq:lfwf2}.
It is then straightforward to express the form factors in our quark-diquark model with the LFWFs. Explicit expressions for proton and neutron are illustrated in Ref.~\cite{ma_flavor_2000}.
For the transverse momentum of the final states, the Drell-Yan-West assignment is employed~\cite{drell_connection_1970, west_phenomenological_1970}, in which
\begin{equation}
  \vec{k}'_{\perp i} = \vec{k}_{\perp i} + (1 - x_i) \vec{q}_\perp,
  \label{eq:dyw_quark}
\end{equation}
for the struck quark, and
\begin{equation}
  \vec{k}'_{\perp j} = \vec{k}_{\perp j} - x_j \vec{q}_\perp,
  \label{eq:dyw_spectator}
\end{equation}
for the spectator.

Experimentally, it is possible to obtain the quark sector contributions to the nucleons by assuming charge symmetry between the $u$ and $d$ quark sectors~\cite{cates_flavor_2011}.
Neglecting the contribution of $s$ quarks, the $u$ and $d$ quark sector form factors for the nucleons are given by
\begin{equation}\label{eq:ff_sector}
F^u_{i} = 2 F^p_i + F^n_i\quad \mathrm{and}\quad F^d_i = F^p_i + 2 F^n_i,
\end{equation}
where $i=1$ or $2$.
The charge symmetry in our model is exact after adopting equal mass and $\beta_{D}$ for the $u$ and $d$ quarks and diquarks composed by them.
This is a natural consequence of the $\mathrm{SU}(6)$ symmetry in our model.
The flavor separation can be achieved by including only one specific quark flavor in the current $J^\mu$.
The result differs from Eqs.~\eqref{eq:ffc_1}-\eqref{eq:ffc_2} in that only LFWFs of the Fock states containing the specific quark are present. For the $u$ sector contribution defined as above, we have
\begin{align}
  F_1^u(Q^2) &= \Braket{\psi_p: P',\uparrow|\frac{\bar{u}\gamma^\mu u}{2P^+}|\psi_p: P,\uparrow} \nonumber\\
  &= \int\frac{\diff^2\vec{k}_\perp \diff x}{16\pi^3} \biggl\{\frac{3}{2} \sum_{i=\{\udarrow\}}
  A^{i,\uparrow\,*}_{uS(ud)}(x,\vec{k}'_\perp)\,A^{i,\uparrow}_{uS(ud)}(x,\vec{k}_\perp) \nonumber\\
  &\quad + \frac{1}{2} \sum_{i=\{\udarrow\}}\,\sum_{j=\{-1,0,1\}}
  B^{ij,\uparrow\,*}_{uV(ud)}(x,\vec{k}'_\perp)\,B^{ij,\uparrow}_{uV(ud)}(x,\vec{k}_\perp)\biggr\}, \label{eq:fu_1}\\
  F_2^u(Q^2) &= -\frac{2M}{q^L} \Braket{\psi_p: P',\uparrow|\frac{\bar{u}\gamma^\mu u}{2P^+}|\psi_p: P,\downarrow} \nonumber\\
  &= -\frac{2M}{q^L} \int\frac{\diff^2\vec{k}_\perp \diff x}{16\pi^3} \biggl\{\frac{3}{2} \sum_{i=\{\udarrow\}}
  A^{i,\uparrow\,*}_{uS(ud)}(x,\vec{k}'_\perp)\,A^{i,\downarrow}_{uS(ud)}(x,\vec{k}_\perp) \nonumber\\
  &\quad + \frac{1}{2} \sum_{i=\{\udarrow\}}\,\sum_{j=\{-1,0,1\}}
  B^{ij,\uparrow\,*}_{uV(ud)}(x,\vec{k}'_\perp)\,B^{ij,\downarrow}_{uV(ud)}(x,\vec{k}_\perp)\biggr\}. \label{eq:fu_2}
\end{align}
The $d$ sector contribution can be obtained in the same way:
\begin{align}
  F_1^d(Q^2) &= \int\frac{\diff^2\vec{k}_\perp \diff x}{16\pi^3} \sum_{i=\{\udarrow\}}\,\sum_{j=\{-1,0,1\}}
  B^{ij,\uparrow\,*}_{dV(uu)}(x,\vec{k}'_\perp)\,B^{ij,\uparrow}_{dV(uu)}(x,\vec{k}_\perp), \label{eq:fd_1}\\
  F_2^d(Q^2) &= -\frac{2M}{q^L}\int\frac{\diff^2\vec{k}_\perp \diff x}{16\pi^3} \sum_{i=\{\udarrow\}}\,\sum_{j=\{-1,0,1\}}
  B^{ij,\uparrow\,*}_{dV(uu)}(x,\vec{k}'_\perp)\,B^{ij,\downarrow}_{dV(uu)}(x,\vec{k}_\perp). \label{eq:fd_2}
\end{align}
The nucleon form factors are combinations of the $u$ and $d$ sector contributions multiplied by charge of the quark according to our definition in Eq.~\eqref{eq:ff_sector}. Form factors for the rest of octet baryons have almost identical expressions to the above except for the coefficients and Fock state components. A similar kind of flavor separation can also be achieved.

\section{Weak transition form factors\label{sec:beta_decay}}
The semileptonic beta decay of baryons can be described by the interaction Hamiltonian in the low energy limit:
\begin{equation}
  H = \frac{G}{\sqrt{2}}J_\mu L^\mu + \mathrm{h.c.},
  \label{eq:weak_hamiltonian}
\end{equation}
where $G$ is the weak coupling constant.
In Eq.~\eqref{eq:weak_hamiltonian}
\begin{equation}
  L^\mu = \bar\psi_e \gamma^\mu(1-\gamma_5)\psi_{\nu_e}
  + \bar\psi_\mu \gamma^\mu(1-\gamma_5)\psi_{\nu_\mu}
  + \bar\psi_\tau \gamma^\mu(1-\gamma_5)\psi_{\nu_\tau}
  \label{eq:lepton_current}
\end{equation}
is the lepton current, and
\begin{equation}
  J^\mu = V^\mu + A^\mu
  \label{eq:hadronic_current}
\end{equation}
is the hadronic current with
\begin{align}
  V^\mu &= V_{ud} \bar{u}\gamma^\mu d + V_{us} \bar{u}\gamma^\mu s, \label{eq:va_current_1}\\
  A^\mu &= - V_{ud} \bar{u}\gamma^\mu\gamma_5 d - V_{us} \bar{u}\gamma^\mu\gamma_5 s, \label{eq:va_current_2}
\end{align}
where $V_{ud}$ and $V_{us}$ are the corresponding CKM matrix elements. The contribution of $\tau$ in the lepton current is usually neglected because of the large $\tau$ mass.
In the limit of zero momentum transfer, the hadronic part of the matrix element can be written as
\begin{align}
  \Braket{P'\lambda' | V^\mu(0) | P\lambda} &= V_{q'q} \bar{u}_{\lambda'}(P')
  \left[ f_1(q^2)\gamma^\mu + i f_2(q^2)\frac{\sigma^{\mu\nu}q_\nu}{M_i}
    + f_3(q^2)\frac{q^\mu}{M_i} \right] u_\lambda(P), \label{eq:me_hadron_1} \\
  \Braket{P'\lambda' | A^\mu(0) | P\lambda} &= V_{q'q} \bar{u}_{\lambda'}(P')
  \left[ g_1(q^2)\gamma^\mu + i g_2(q^2)\frac{\sigma^{\mu\nu}q_\nu}{M_i}
    + g_3(q^2) \frac{q^\mu}{M_i} \right]\gamma_5 u_\lambda(P), \label{eq:me_hadron_2}
\end{align}
where $q = p_i-p_f$, and $M_i$ is the mass of the initial baryon. The term $f_1$, $f_2$ and $f_3$ stand for the vector, induced tensor (``weak magnetism'') and induced scalar form factors, while $g_1$, $g_2$ and $g_3$ are the axial vector, induced pseudotensor (``weak electricity'') and induced pseudoscalar form factors.

After adopting the Drell-Yan frame as in Eqs.~\eqref{eq:dy_frame_1}-\eqref{eq:dy_frame_2}, the form factors can be extracted as
\begin{align}
  f_1(q^2) &= \Braket{P', \uparrow | \frac{V^+(0)}{2P^+} | P, \uparrow}, \label{eq:ffe_1}\\
  -q^L f_2(q^2) &= M_i \Braket{P', \uparrow | \frac{V^+(0)}{2P^+} | P, \downarrow}, \label{eq:ffe_2}\\
  g_1(q^2) &= \Braket{P', \uparrow | \frac{A^+(0)}{2P^+} | P, \uparrow}, \label{eq:ffe_3}\\
  q^L g_2(q^2) &= M_i \Braket{P', \uparrow | \frac{A^+(0)}{2P^+} | P, \downarrow}. \label{eq:ffe_4}
\end{align}
$f_3$ and $g_3$ are not calculated since the last terms in the two equations of Eqs.~\eqref{eq:me_hadron_1}-\eqref{eq:me_hadron_2} vanish after setting $q^+$ to $0$.
Eqs.~\eqref{eq:ffe_1}-\eqref{eq:ffe_4} resembles Eqs.~\eqref{eq:ff_plus_1}-\eqref{eq:ff_plus_2} much although it should be noted that the initial and final state in Eqs.~\eqref{eq:ffe_1}-\eqref{eq:ffe_4} are different. It is straightforward to write these form factors in the form of LFWFs as in Eqs.~\eqref{eq:fu_1}-\eqref{eq:fu_2} and Eqs.~\eqref{eq:fd_1}-\eqref{eq:fd_2}.

\section{Numerical results\label{sec:numerical}}
We apply different forms of momentum space wave functions as described by Eq.~\eqref{eq:mswf_bhl} and Eq.~\eqref{eq:mswf_pl} in our numerical calculation.
The parameters used are given in Tab.~\ref{tab:par_model}.
We present our numerical results in Tab.~\ref{tab:results}.

Parameters in Set~I and Set~II are used with the BHL wave function. Set~I is chosen according to Ref.~\cite{ma_flavor_2000} as a first-attempt.
The quark masses are identical to those used in non-relativistic models.
The $\beta_D$ parameter, which is presumed to be of same magnitude of the quark mass, acts as a characterization of the transverse momentum of quark and diquark.
In Set~I both $\beta_S$ and $\beta_V$ are simply set to be the same as $m_u$.
It is shown in Tab.~\ref{tab:results} that this set of relatively crude parameters is able to give a reasonable result, although the magnetic moments are noticeably smaller than experimental results.
This is partly because our choice of quark masses is slightly larger than needed, which leads to a suppression of the form factors.

\begin{figure*}
  \subfloat[\label{proton_ge}]{%
    \includegraphics[width=0.49\textwidth]{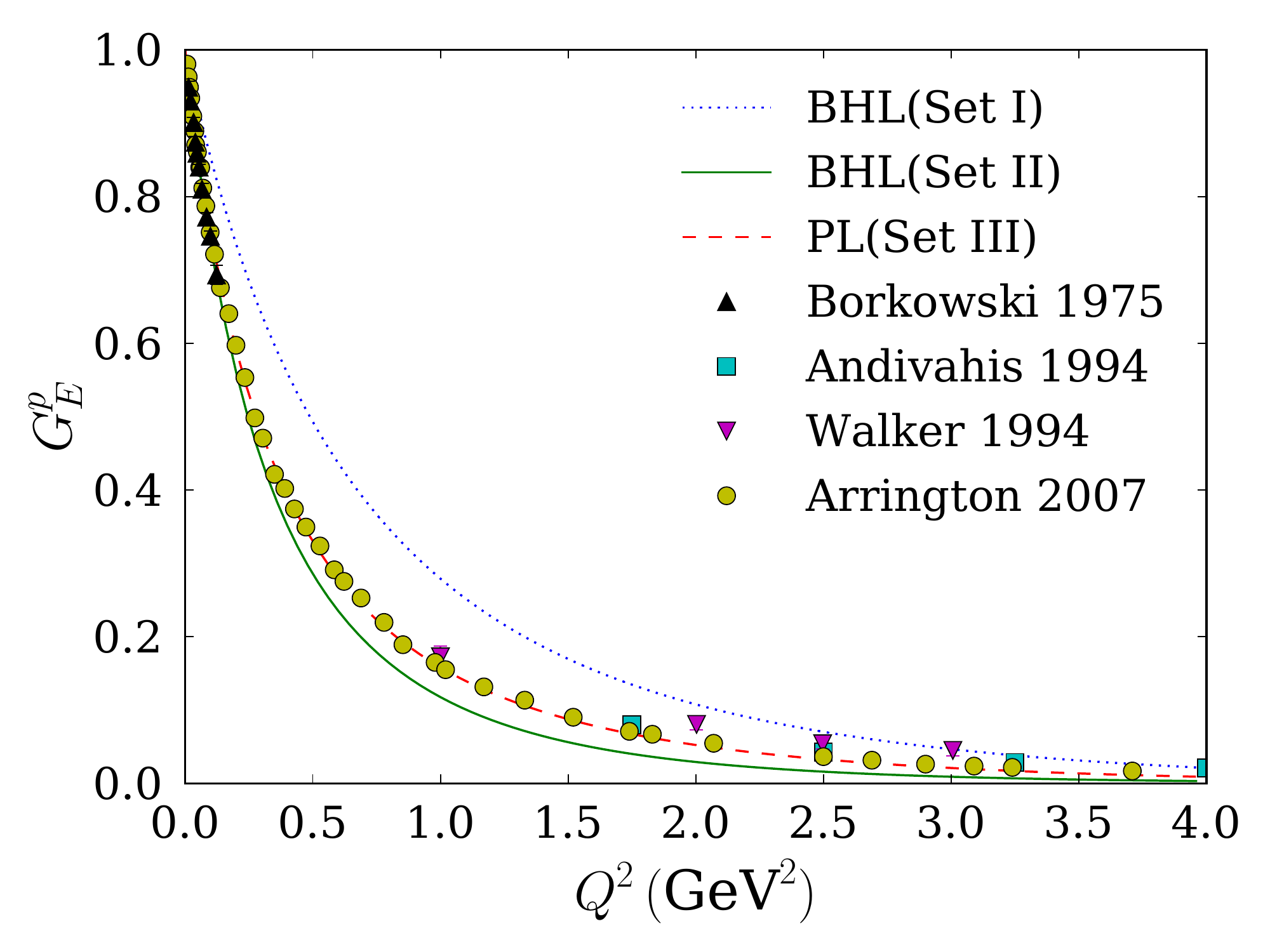}%
  }\hfill
  \subfloat[\label{proton_gm}]{%
    \includegraphics[width=0.49\textwidth]{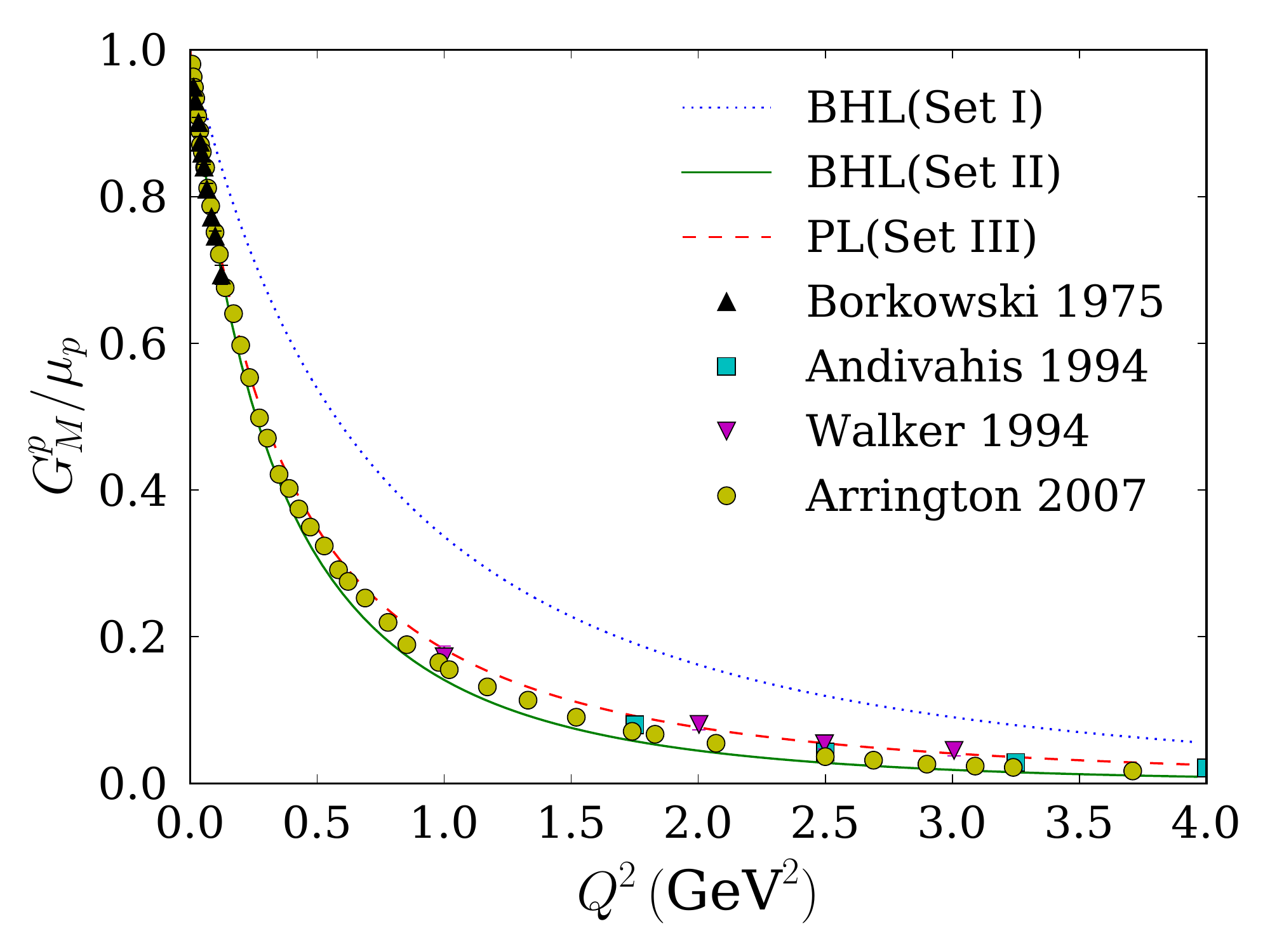}%
  }
  \caption{(Color online) Electric~\protect\subref{proton_ge} and magnetic~\protect\subref{proton_gm} form factors of proton calculated with different wave functions and parameters. Data are from Ref.~\cite{borkowski_electromagnetic_1975,*andivahis_measurements_1994,*walker_measurements_1994,*arrington_global_2007}\label{fig:ffproton}}
\end{figure*}

\begin{figure*}
  \subfloat[\label{neutron_ge}]{%
    \includegraphics[width=0.49\textwidth]{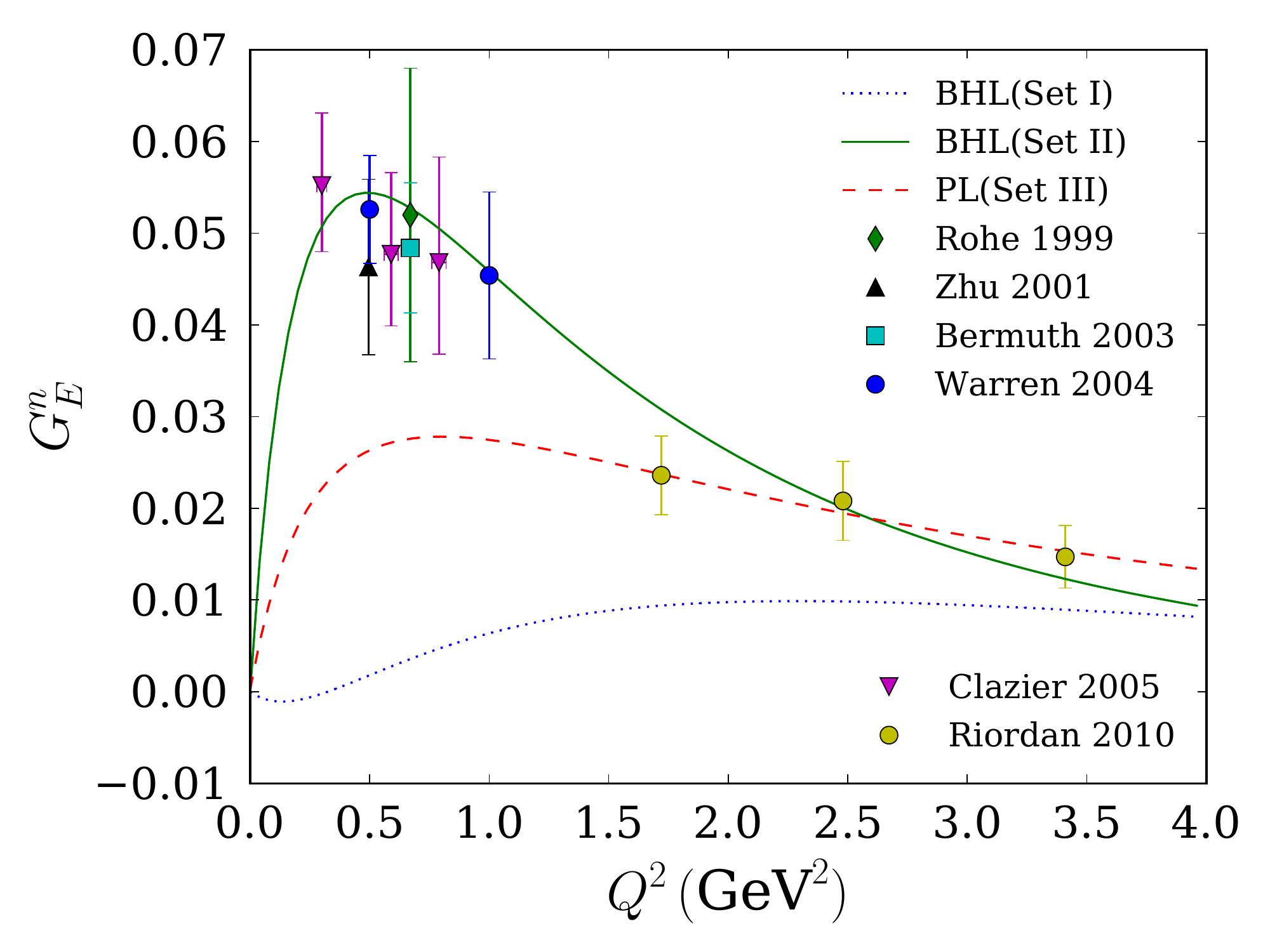}%
  }\hfill
  \subfloat[\label{neutron_gm}]{%
    \includegraphics[width=0.49\textwidth]{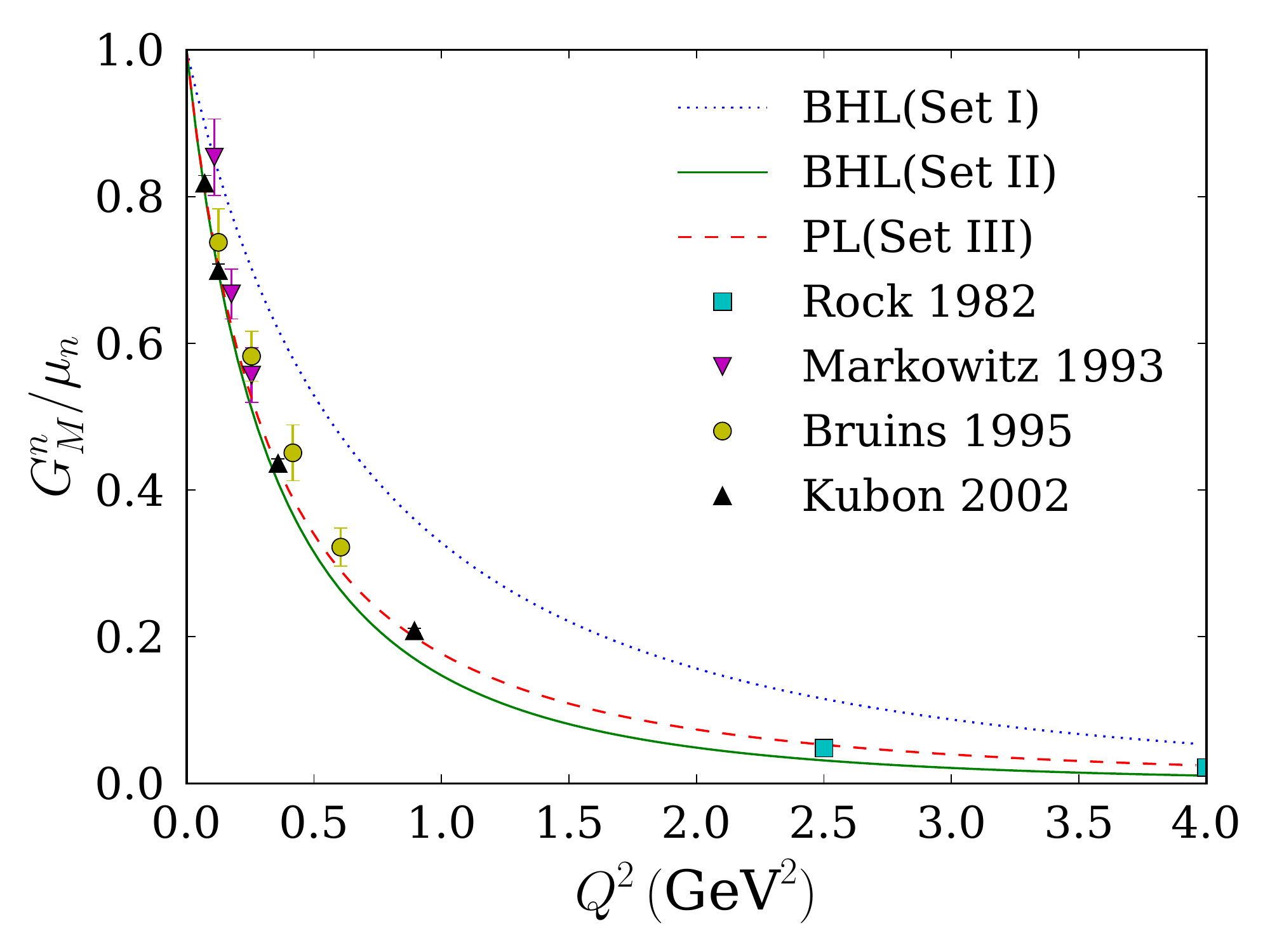}%
  }
  \caption{(Color online) Electric~\protect\subref{neutron_ge} and magnetic~\protect\subref{neutron_gm} form factors of neutron calculated with different wave functions and parameters. Data are from Ref.~\cite{rohe_measurement_1999,*zhu_measurement_2001,*bermuth_neutron_2003,*warren_measurement_2004,*glazier_measurement_2005,*riordan_measurements_2010,*rock_measurement_1982,*markowitz_measurement_1993,*bruins_measurement_1995,*kubon_precise_2002}\label{fig:ffneutron}}
\end{figure*}

\begin{figure*}
  \subfloat[]{%
    \includegraphics[width=0.49\textwidth]{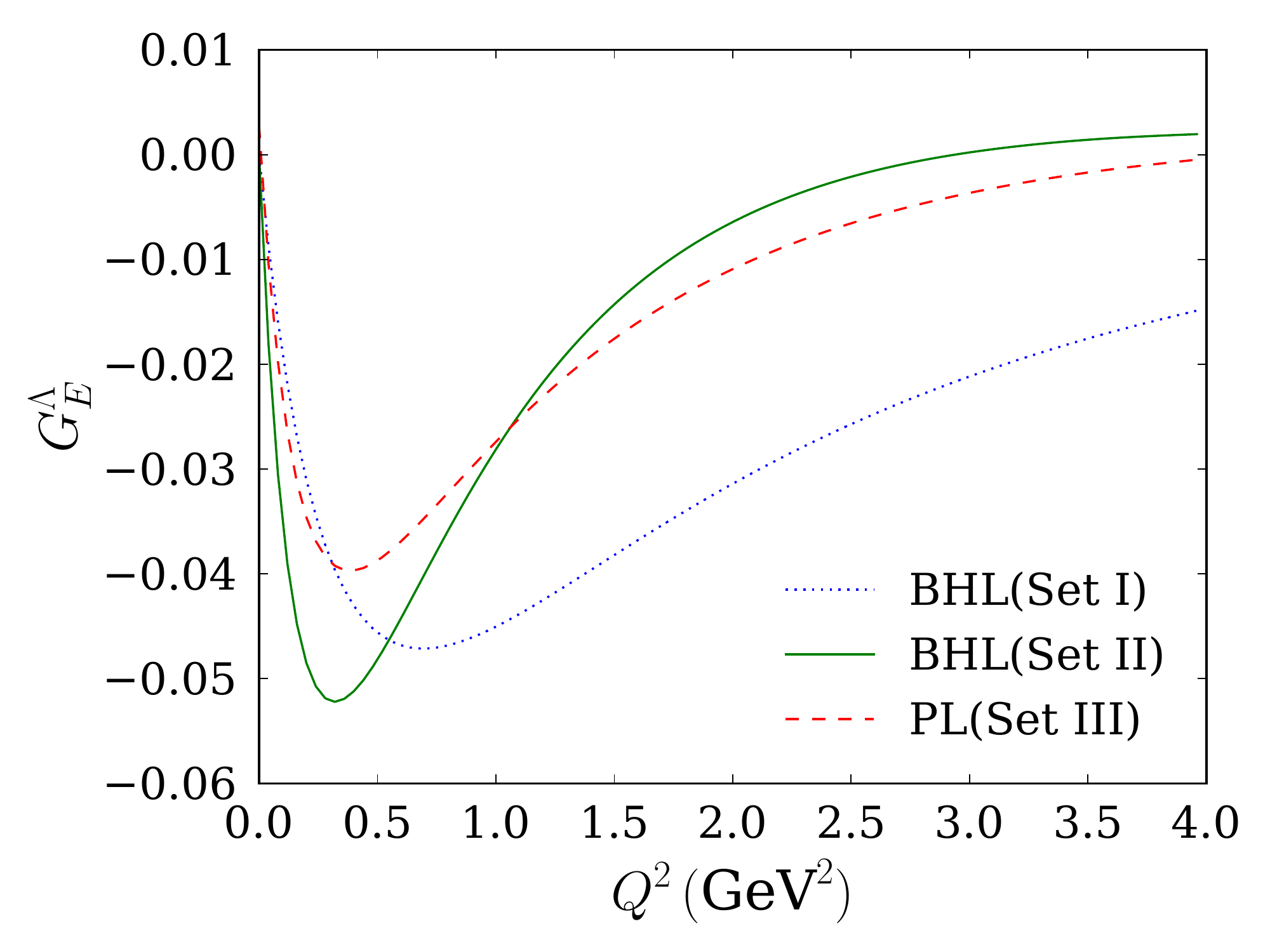}%
  }\hfill
  \subfloat[]{%
    \includegraphics[width=0.49\textwidth]{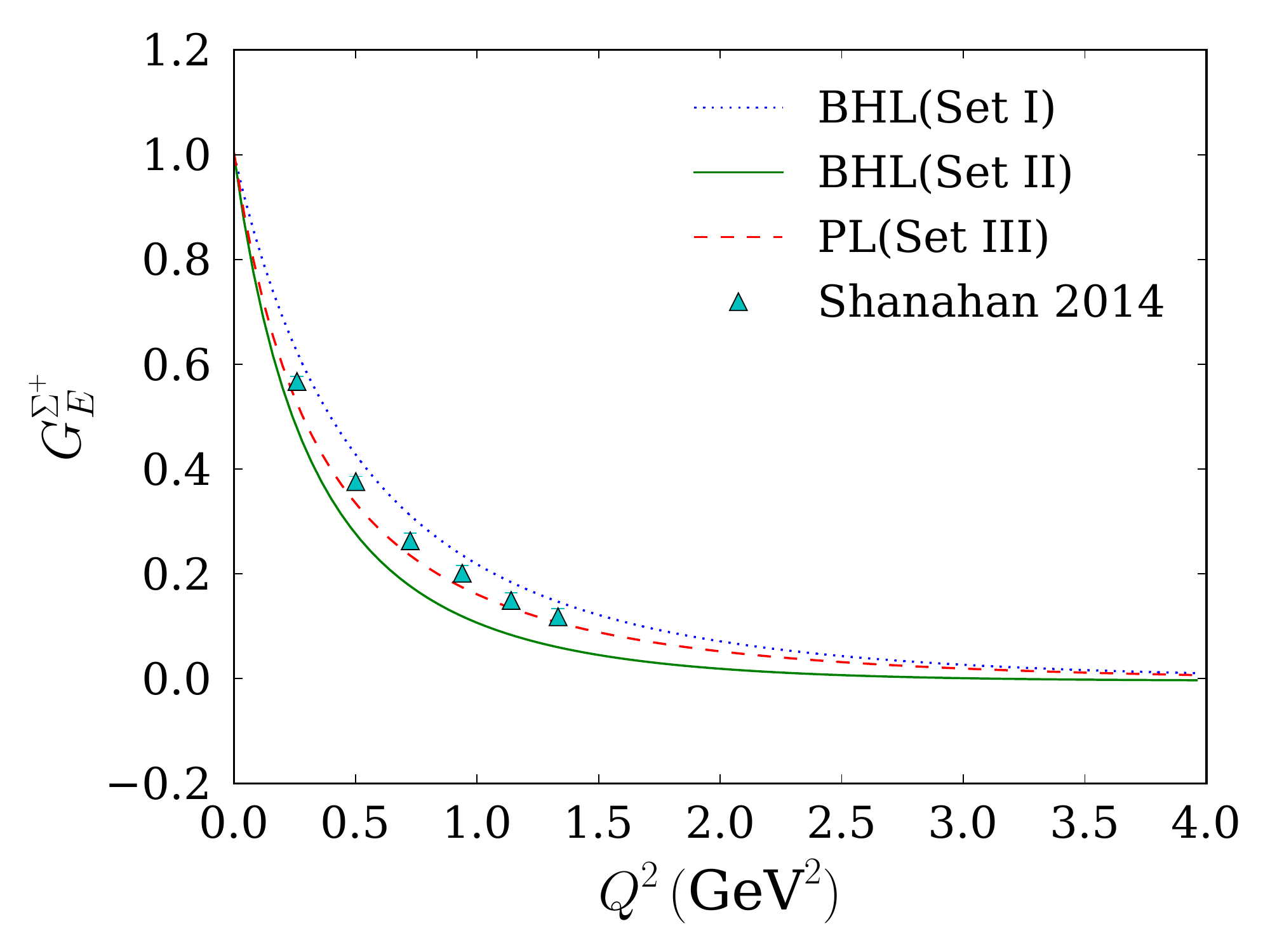}%
  }\\
  \subfloat[]{%
    \includegraphics[width=0.49\textwidth]{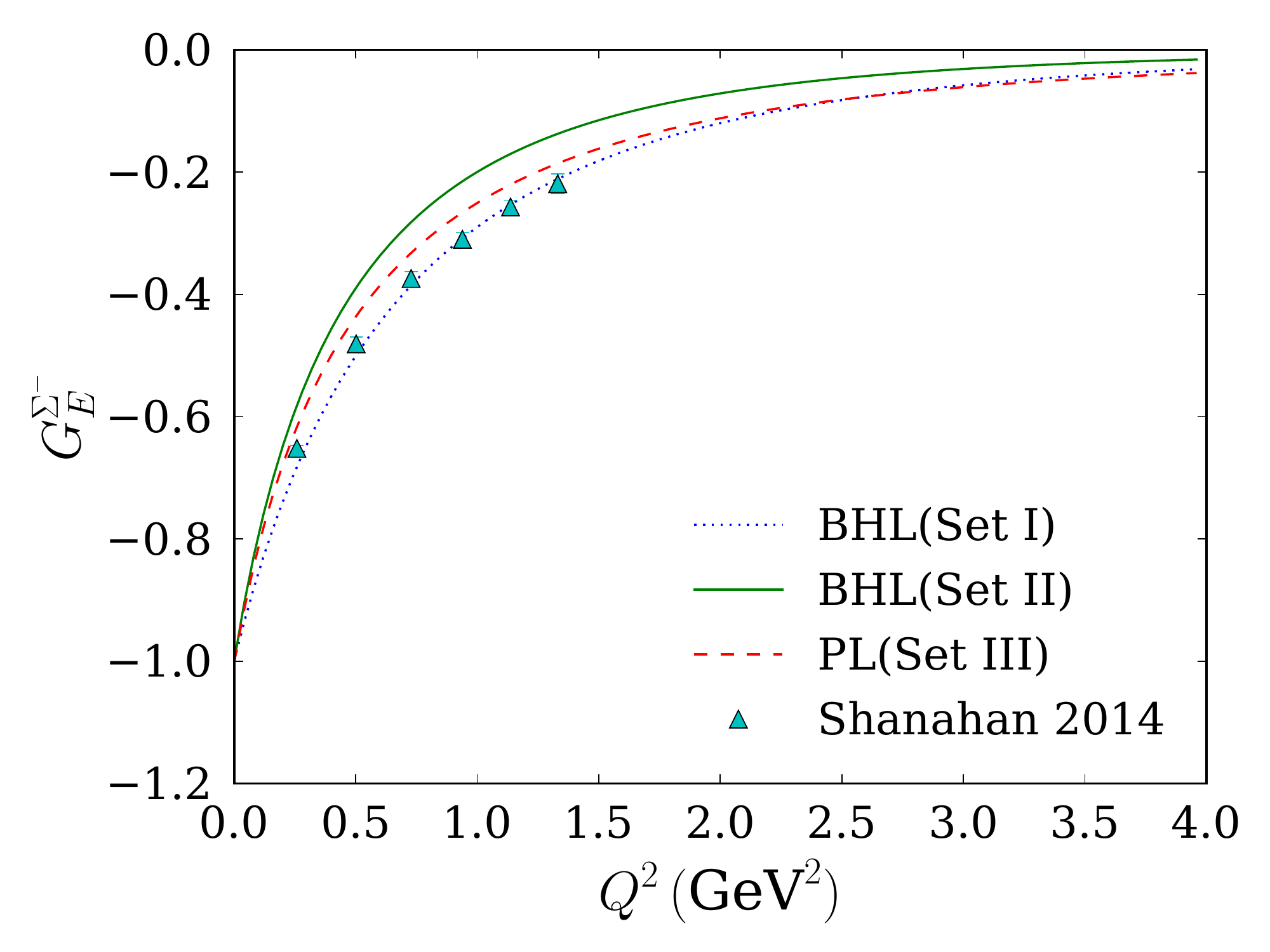}%
  }\hfill
  \subfloat[]{%
    \includegraphics[width=0.49\textwidth]{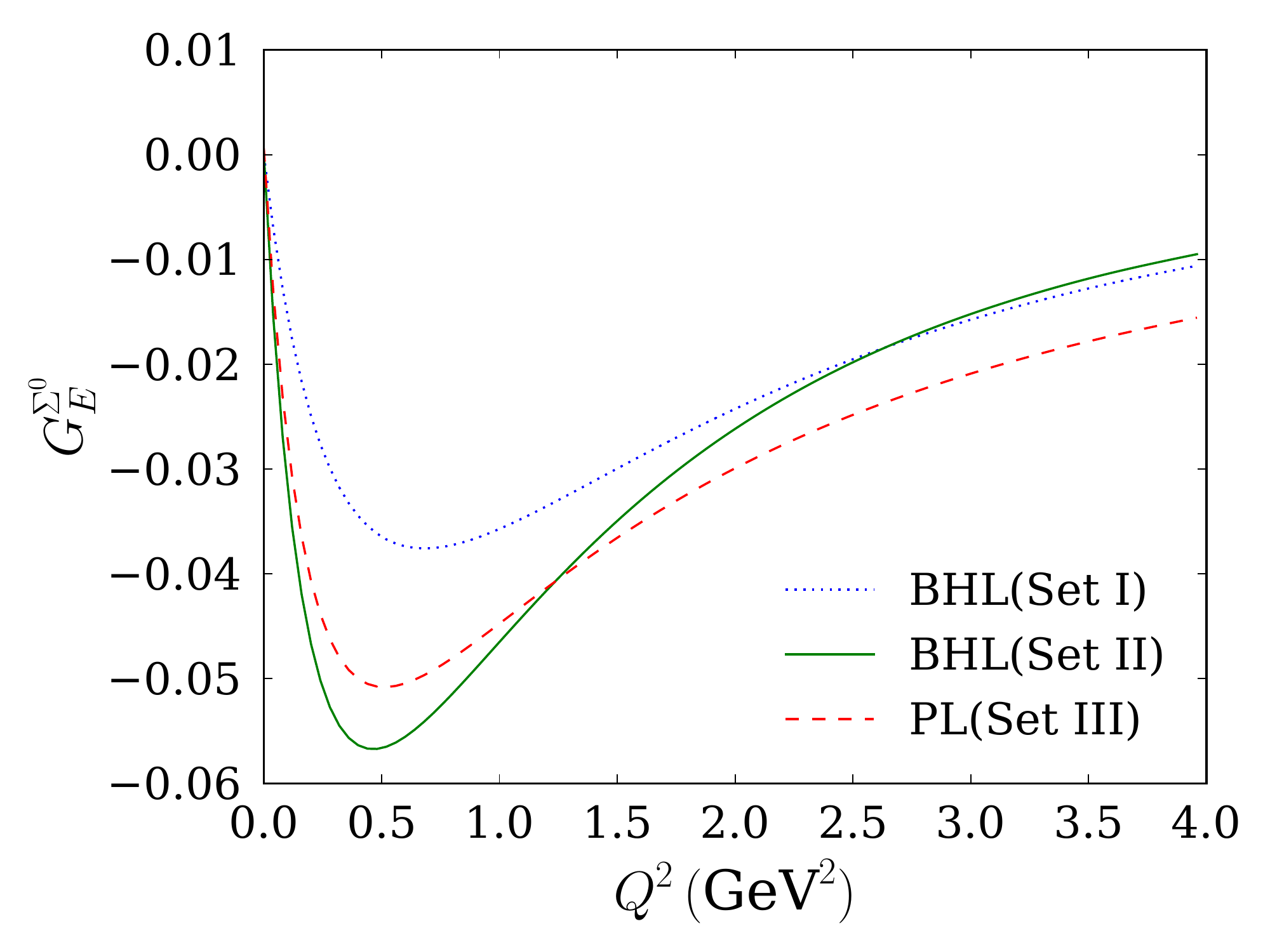}%
  }\\
  \subfloat[]{%
    \includegraphics[width=0.49\textwidth]{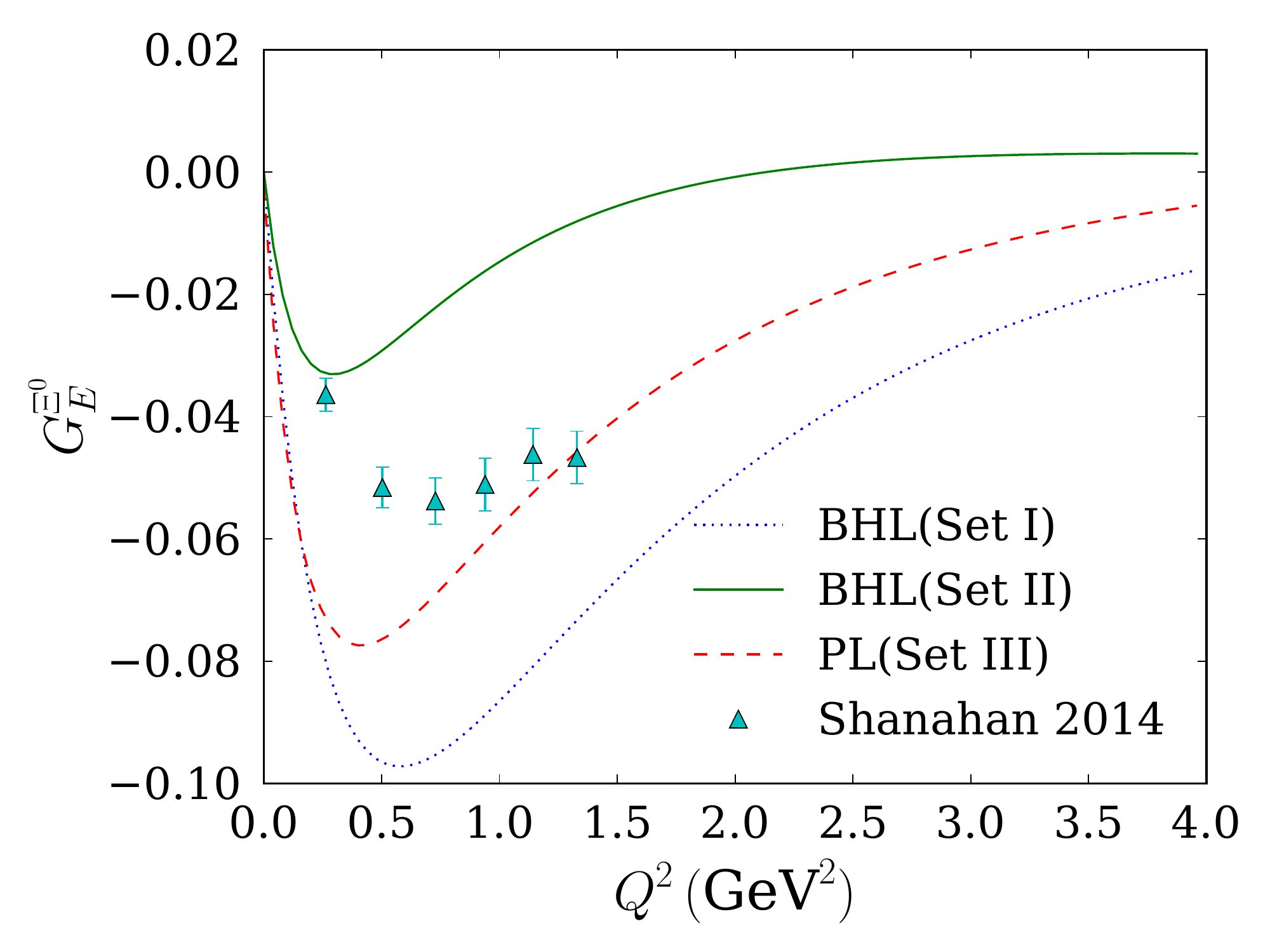}%
  }\hfill
  \subfloat[]{%
    \includegraphics[width=0.49\textwidth]{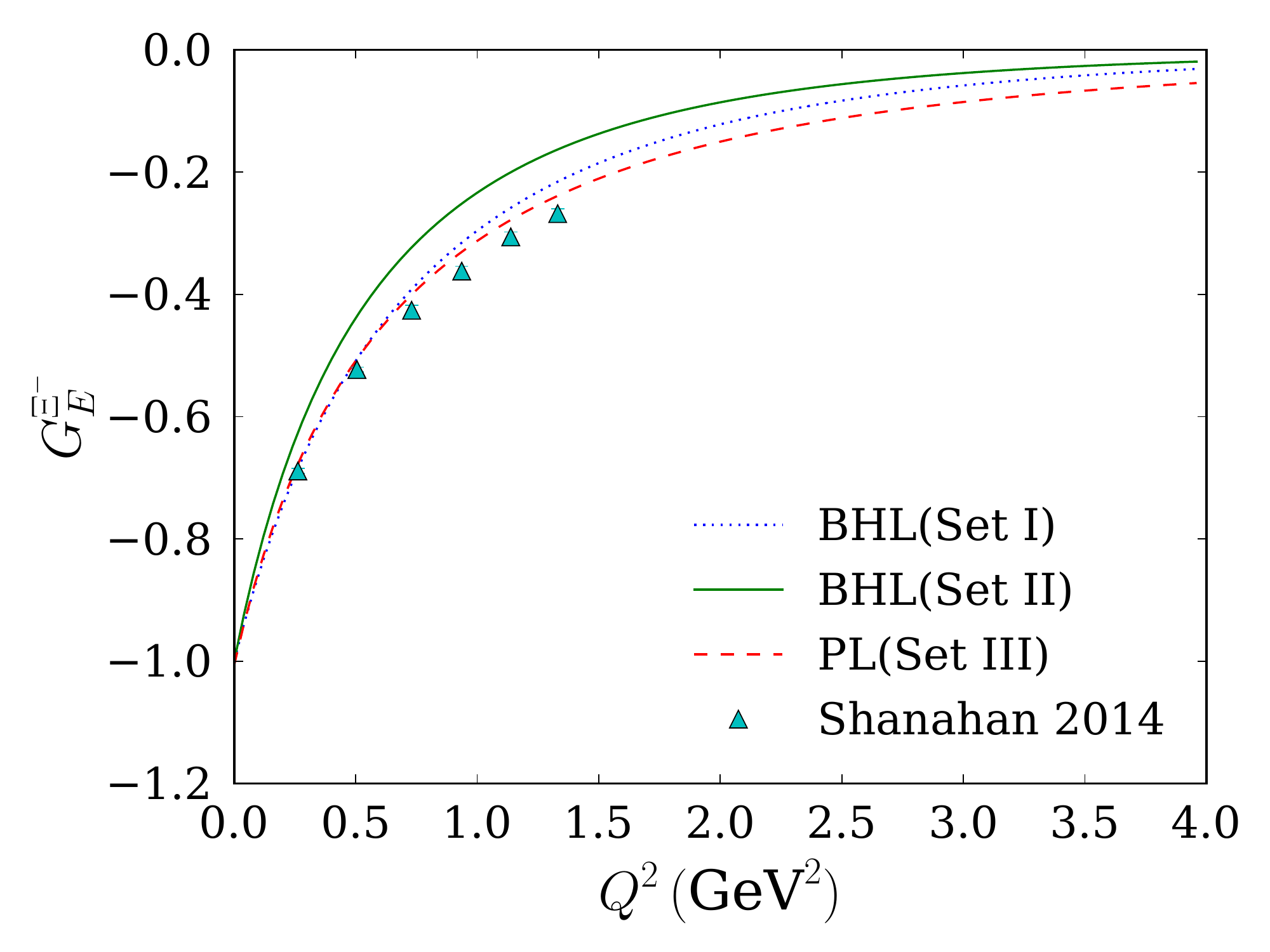}%
  }
  \caption{(Color online) Electric form factor of the octet baryons calculated with different wave functions and parameters, compared with lattice results in Ref.~\cite{shanahan_electric_2014}.\label{fig:gebaryons}}
\end{figure*}

\begin{figure*}
  \subfloat[]{%
    \includegraphics[width=0.49\textwidth]{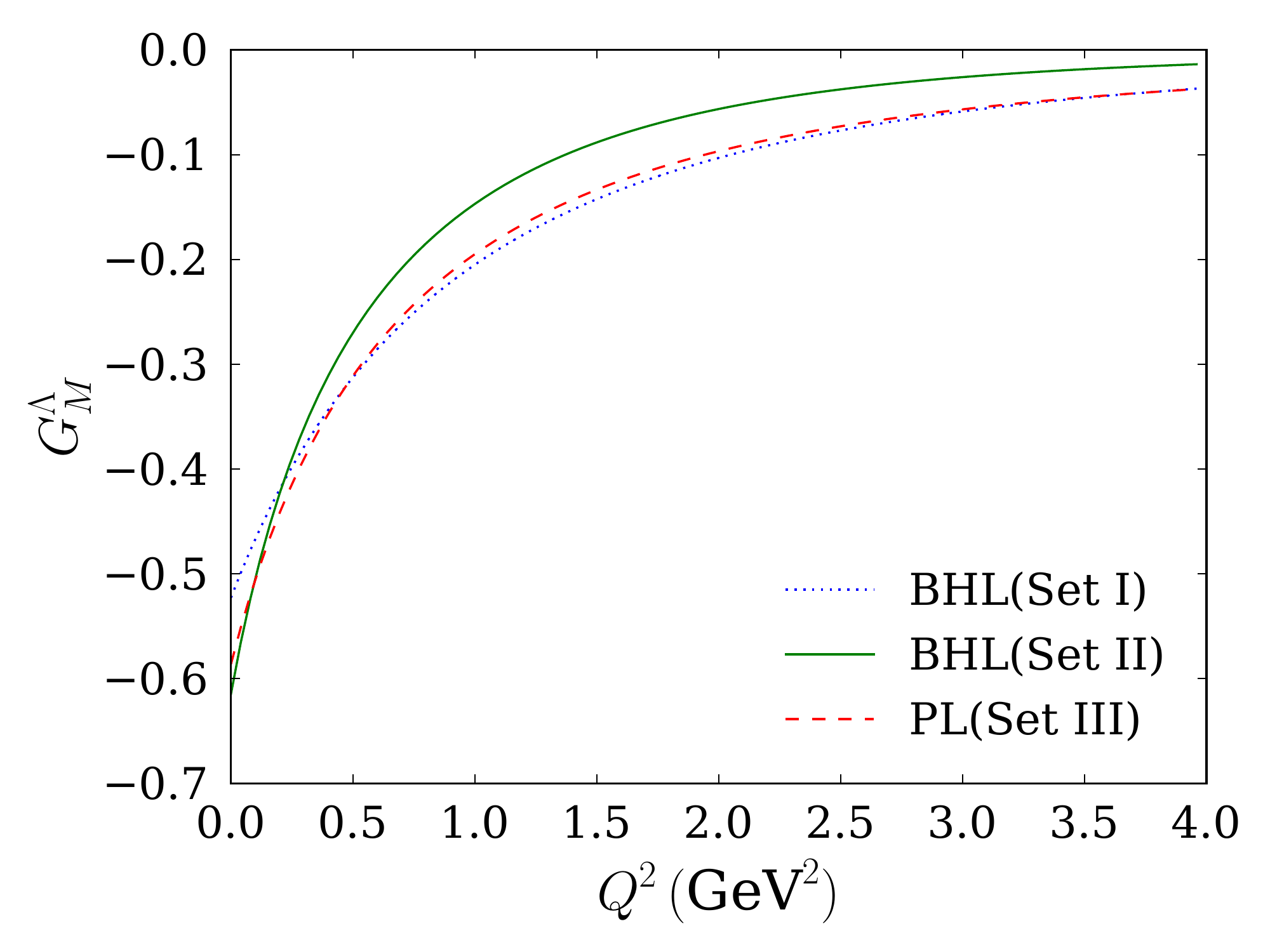}%
  }\hfill
  \subfloat[]{%
    \includegraphics[width=0.49\textwidth]{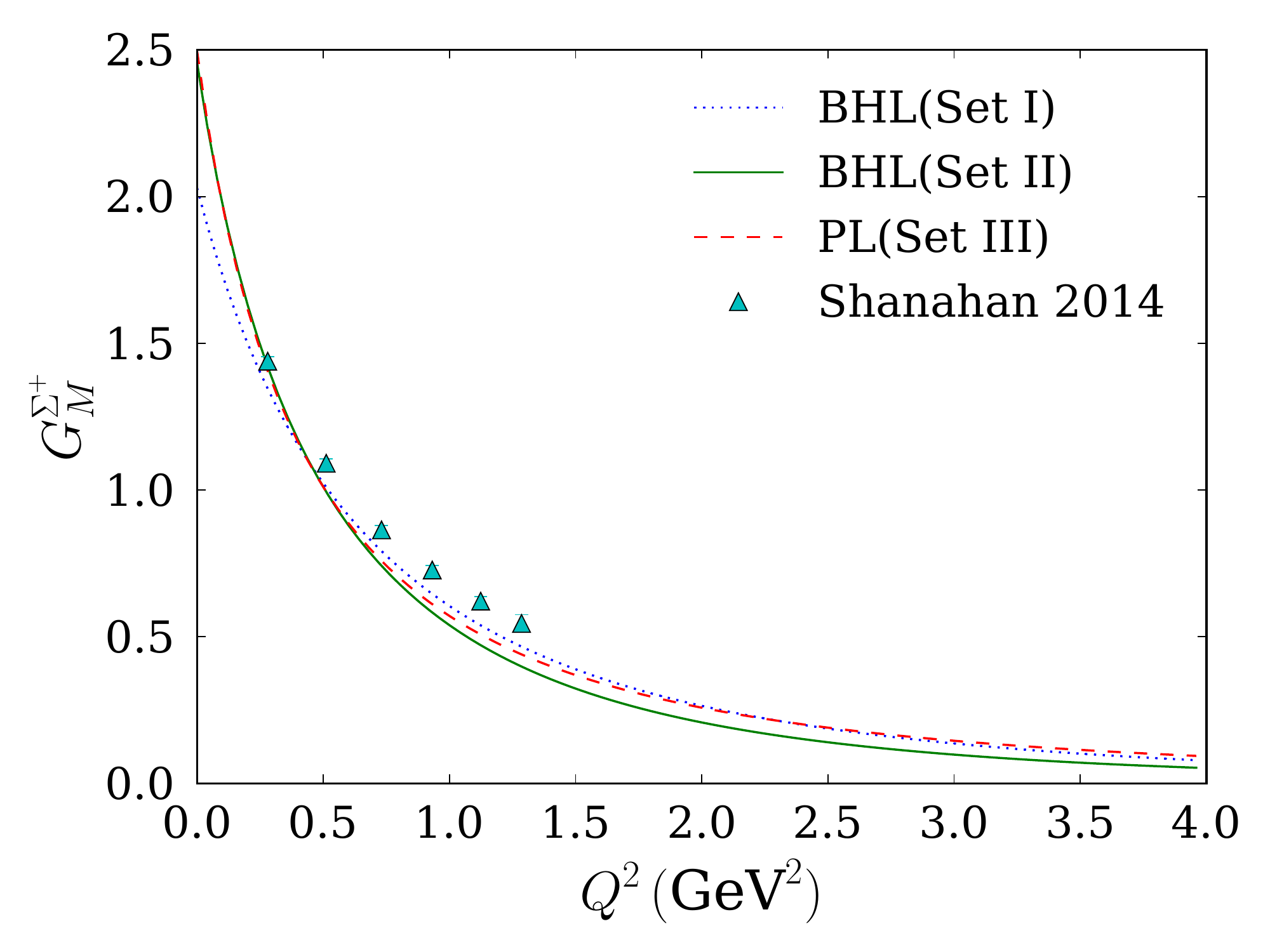}%
  }\\
  \subfloat[]{%
    \includegraphics[width=0.49\textwidth]{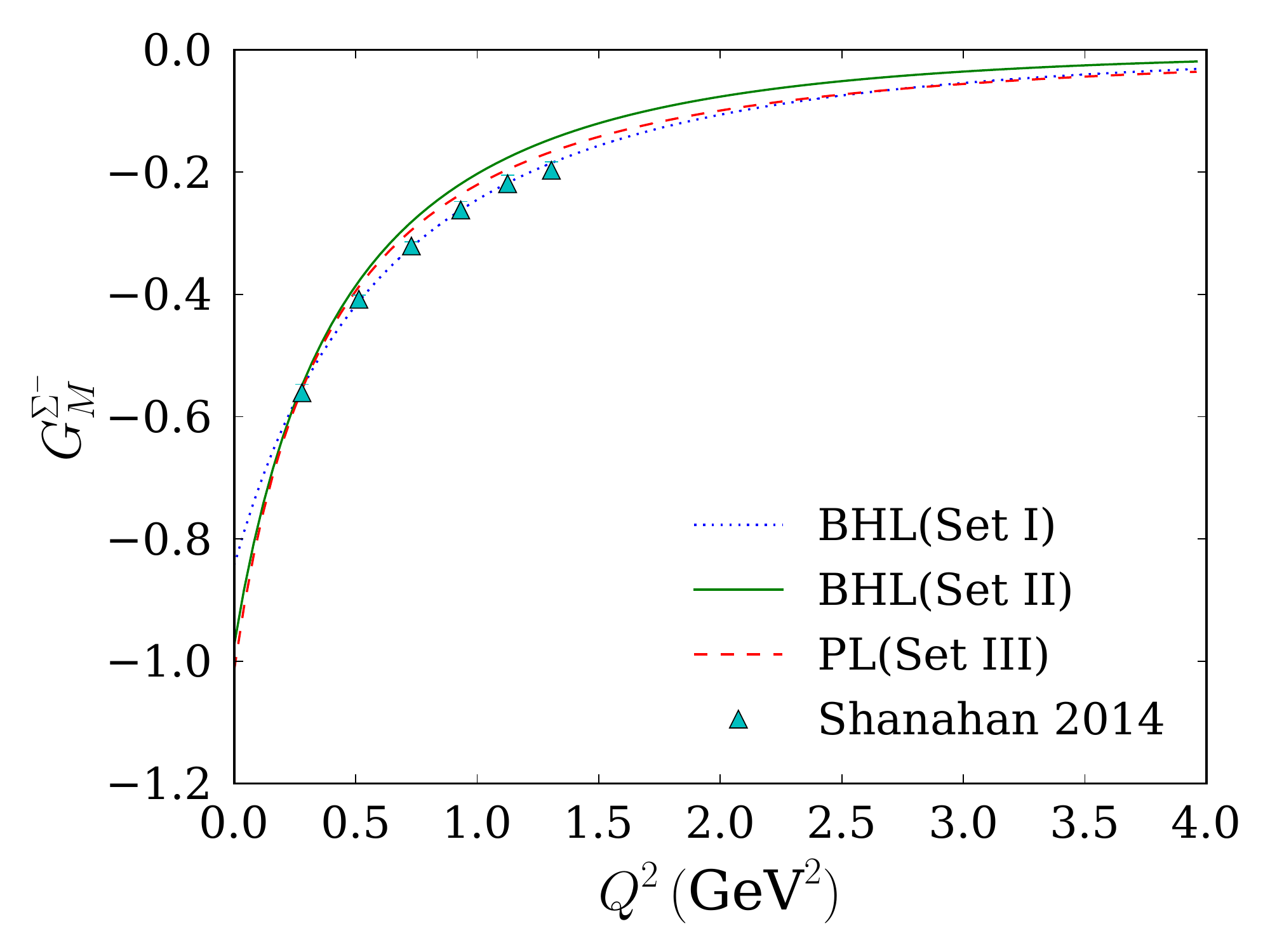}%
  }\hfill
  \subfloat[]{%
    \includegraphics[width=0.49\textwidth]{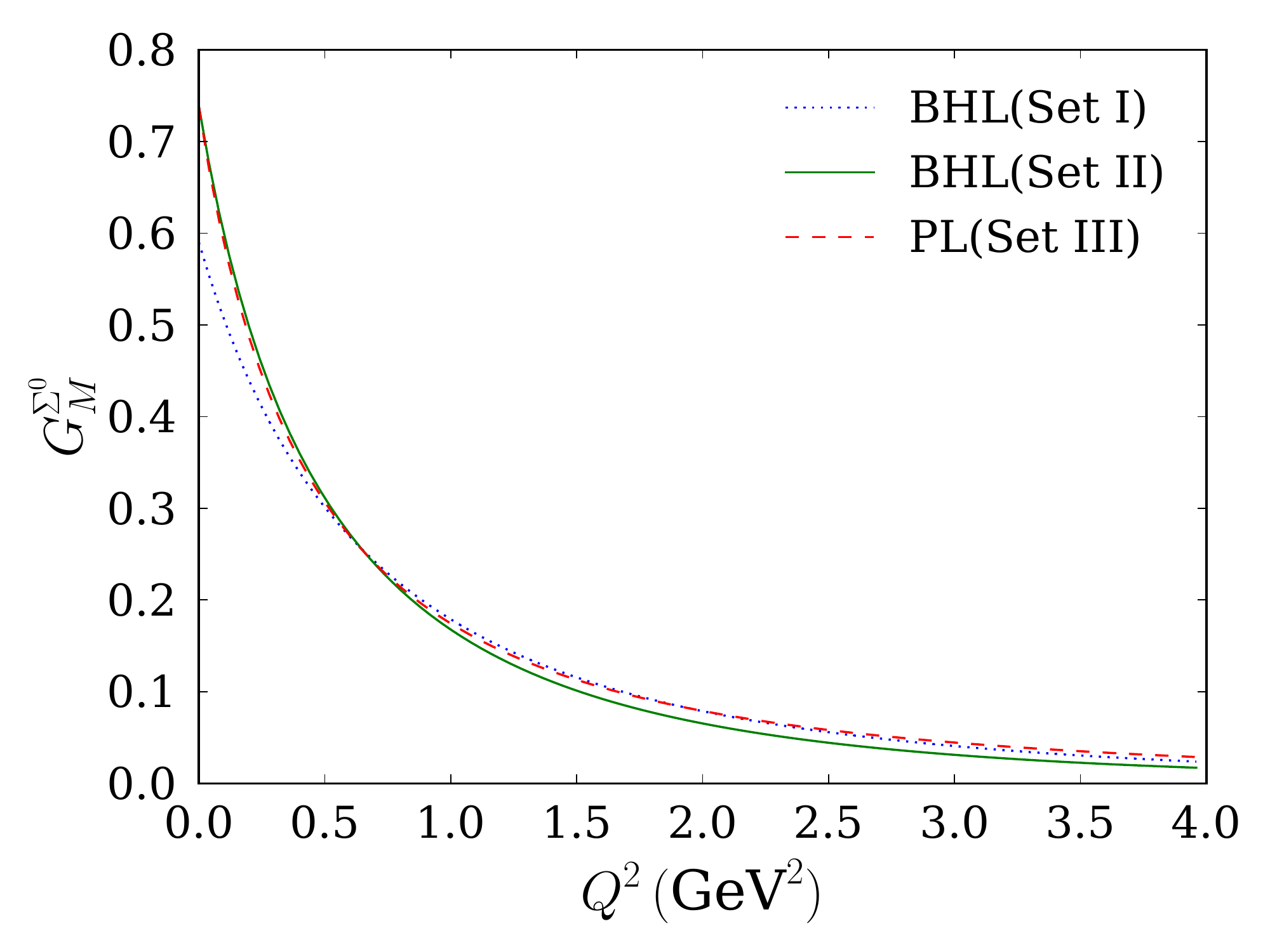}%
  }\\
  \subfloat[]{%
    \includegraphics[width=0.49\textwidth]{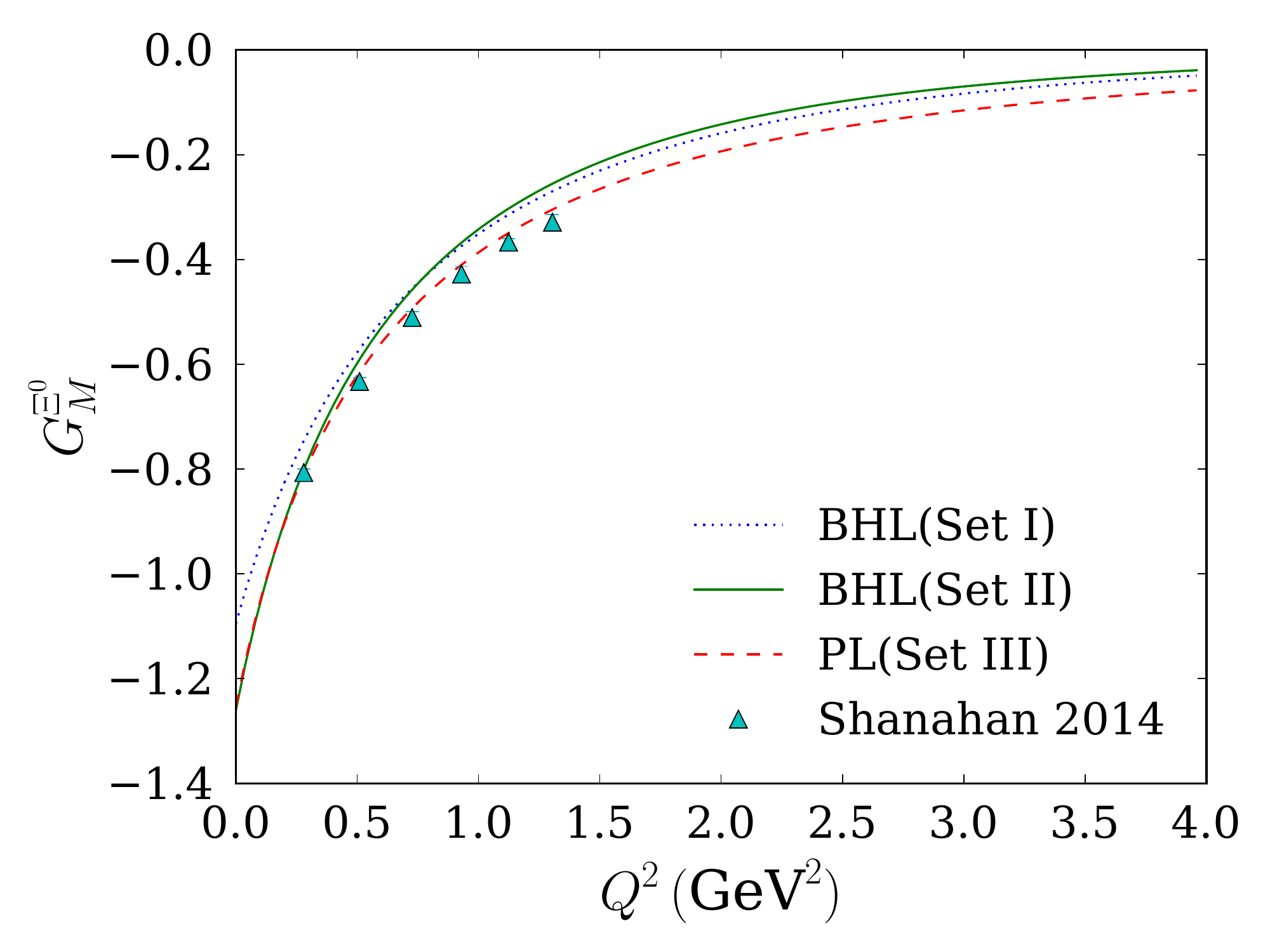}%
  }\hfill
  \subfloat[]{%
    \includegraphics[width=0.49\textwidth]{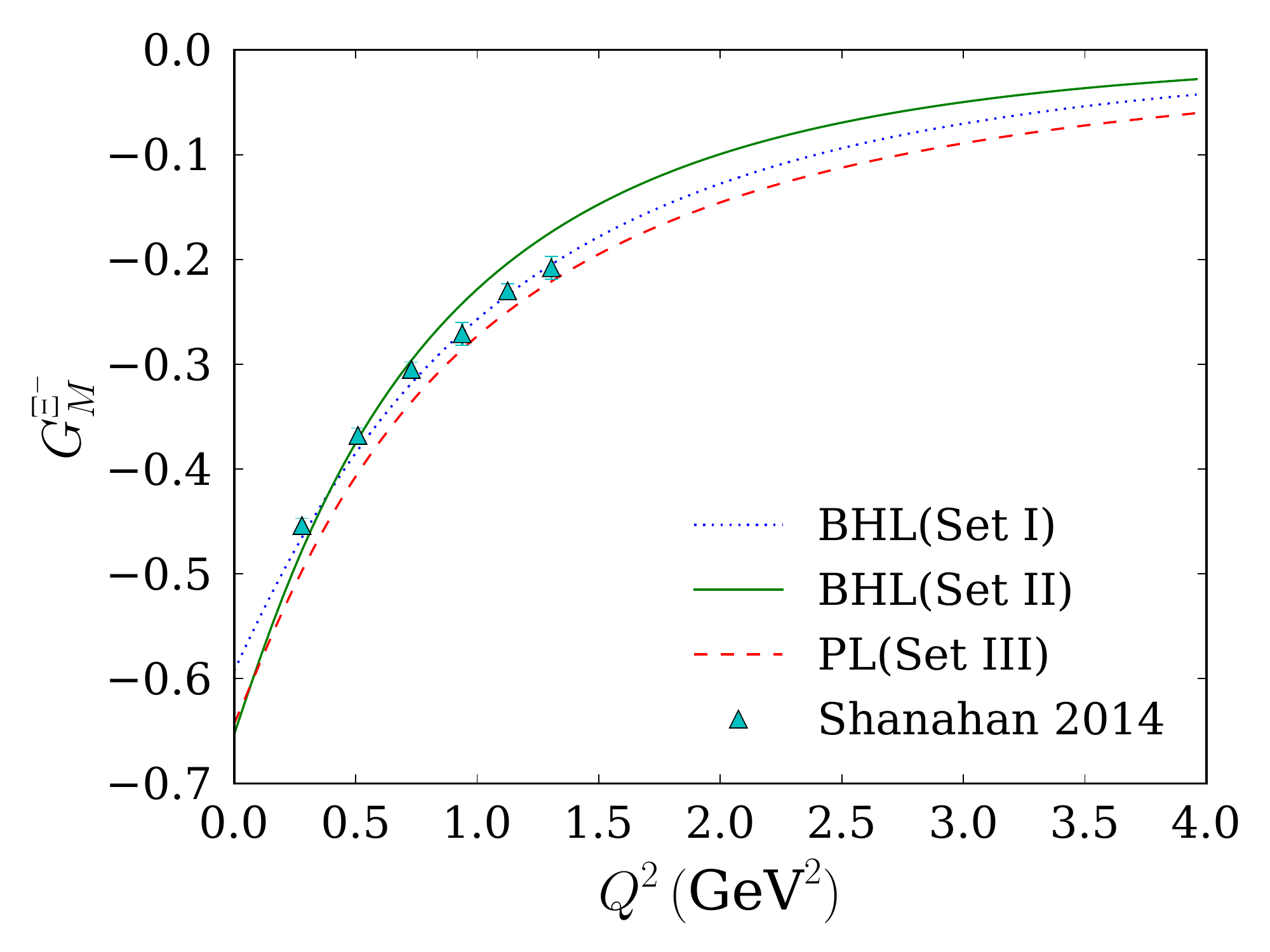}%
  }
  \caption{(Color online) Magnetic form factor of the octet baryons calculated with different wave functions and parameters, compared with lattice results in Ref.\cite{shanahan_magnetic_2014}.\label{fig:gmbaryons}}
\end{figure*}

Parameters in Set~II give  better results overall for both magnetic moments and weak charges.
The quark masses are identical to a previous work~\cite{ma_electromagnetic_2002} and other relativistic quark models~\cite{schlumpf_relativistic_1994}.
As has been pointed out in Ref.~\cite{ma_electromagnetic_2002}, the difference of $\beta_D$ between the scalar and axial-vector diquark states is necessary in order to reproduce the magnetic moments and the electric and magnetic radius.
There are abundant experimental results for the nucleon form factors form unpolarized measurements and more recent double polarization measurements.
The calculated electromagnetic form factors of the nucleons as functions of $Q^2$ are shown in Fig.~\ref{fig:ffproton} and Fig.~\ref{fig:ffneutron} up to $Q^2=4\,\mathrm{GeV}^2$ and compared with experimental measurements.
The form factors for the other octet members are presented in Fig.~\ref{fig:gebaryons} and Fig.~\ref{fig:gmbaryons}.
These form factors have been under some theoretical studies although no measurement has been made so far.
Our result is compared with Refs.~\cite{shanahan_magnetic_2014, shanahan_electric_2014} which is obtained from lattice QCD with a chiral extrapolation.
There is a noticeable deviation between the two results. 
Namely our results decrease a bit more rapidly with $Q^2$.
As has been pointed out in Ref.~\cite{shanahan_magnetic_2014}, the lattice results need a ``slightly greater curvature in the $Q^2$ fit functions'' to achieve a better agreement with experiment, which means our model results might be giving just the right trend of the curves.

We also employ the power-law form wave function Eq.~\eqref{eq:mswf_pl} in the calculation of magnetic moments.
Parameters are given as Set~III in Tab.~\ref{tab:par_model}.
As the BHL prescription features a Gaussian distribution around the invariant mass, the exponential falloff is too strong for higher $Q^2$.
Thus the numerical results of the BHL prescription should only be valid in the low and moderate $Q^2$ region up to $4\,\GeV^2$.
At high $Q^2$, the power-law form can be a better choice~\cite{brodsky_wave_1994}.
For the results of magnetic moments and axial couplings, the variation resulted from different wave functions is relatively small.
This fact suggests that the qualitative properties at low $Q^2$ are essentially determined by the flavor and spin structures.
The detailed form of the momentum space wave functions only affects our results in a quantitative way.
The same behavior is also present in the model calculation of Ref.~\cite{schlumpf_relativistic_1994}.

\begin{figure*}
  \subfloat[\label{ff_flavor_f1}]{%
    \includegraphics[width=0.49\textwidth]{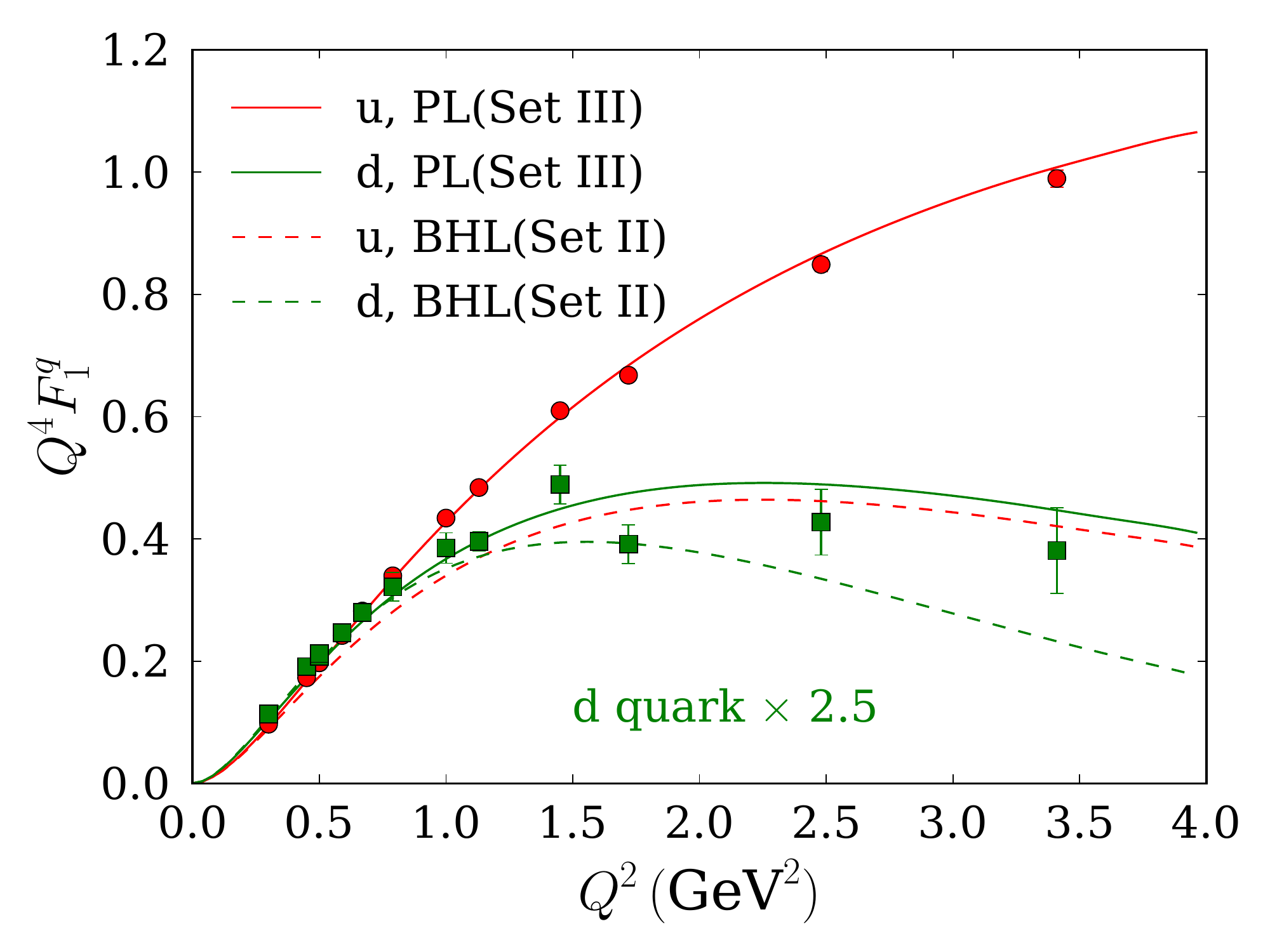}%
  }\hfill
  \subfloat[\label{ff_flavor_f2}]{%
    \includegraphics[width=0.49\textwidth]{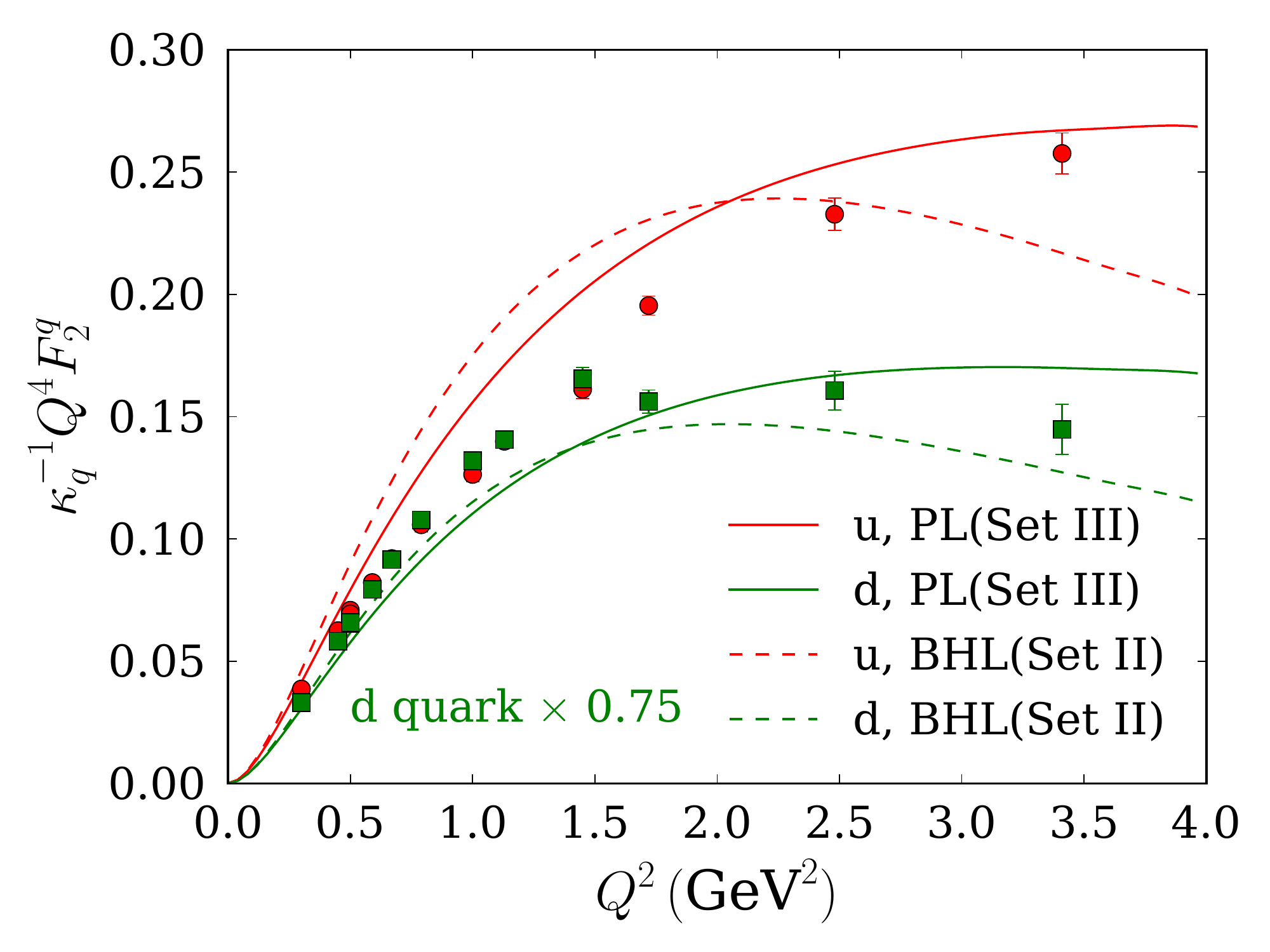}%
  }
  \caption{(Color online) The $u$- and $d$-contributions to the proton form factors (multiplied by $Q^4$). The $d$-contributions are multiplied by a factor for better comparison. The normalization factor $\kappa_q$ is given by $\kappa_q=F_2^q(0)$. Experimental data are from Ref.~\cite{cates_flavor_2011}.\label{fig:udneutron}}
\end{figure*}

Experimentally, the flavor separation of the nucleon electromagnetic form factor has been achieved by combining data of corresponding Sachs form factors~\cite{cates_flavor_2011}. 
The flavor separated form factors in our model are obtained with Eqs.~\eqref{eq:fu_1}--\eqref{eq:fd_2} and compared with experimental results in Fig.~\ref{fig:udneutron}. 
Up to $Q^2=1 \GeV^2$ the $u$- and $d$-contributions are identical after scaling the d-contribution to $F_1$ and $F_2$ with a factor of 2.5 and 0.75 respectively. 
Above $1\,\GeV^2$ the two contribution differ from each other. 
In our model this difference comes from the $\mathrm{SU}(6)$ spin-flavor structure and the Melosh-Wigner rotation. 
The power-law wave function gives a reasonable description of both the contribution to $F_1$ and $F_2$, while the BHL wave function produces significantly lower results for the $F_1$ case. 
The strong fall-off of the BHL results in the high $Q^2$ region is partially a result of the exponential in the wave function, suggesting again that the power-law wave function could be a better choice for higher $Q^2$.
Flavor separated contributions to the octet magnetic moments can also be obtained using Eqs.~\eqref{eq:fu_1}-\eqref{eq:fd_2} for proton and their variants for other baryons. The results are presented in Tab.~\ref{tab:magnetic_moments_flavor}.

The vector and axial-vector couplings $f_1(0)$ and $g_1(0)$ for the semileptonic decays are listed in Tab.~\ref{tab:coupling} for completeness.
Our results of the couplings are in agreement with another light-cone model calculation in Ref.~\cite{schlumpf_relativistic_1994}.
In the $\mathrm{SU}(3)$ symmetry limit, the weak electric form factor $g_2$ should vanish.
Our model indeed gives vanishing or very small $g_2$ despite the fact that different mass and $\beta$ have been used for the quarks and diquarks which breaks the $\mathrm{SU}(3)$ symmetry explicitly.
For all the decays calculated, our model gives $|g_2(0)/f_1(0)|<0.2$. As $f_1(0)$ vanishes for $\Sigma^- \to \Lambda e^- \bar{\nu}_e$, $g_2$ is compared in this case with $g_1$ instead.

Additionally, the spin contents of proton can serve as another test of our model. The quark spin content of proton $\Delta \Sigma$ is given by
\begin{equation}
  \Delta \Sigma = \Delta u + \Delta d,
\end{equation}
where $\Delta u$ and $\Delta d$ are the first moments of the helicity distribution functions of $u$ and $d$ quark. 
For a proton, given the spin-flavor wave function in Eqs.~\eqref{eq:plfwf_1}-\eqref{eq:plfwf_3}, the quark helicity distributions are given as
\begin{align}
  \Delta u(x, \vec{k}_\perp) =&\ \frac{3}{2} \left( \lvert A^{\uuarrow}_{uS(ud)} (x, \vec{k}_\perp) \rvert^2 - \lvert A^{\duarrow}_{uS(ud)} (x, \vec{k}_\perp) \rvert^2 \right) \nonumber\\
  &+ \frac{1}{2} \sum_{j=\{-1,0,1\}} \left(\lvert B^{\uparrow j, \uparrow}_{uV(ud)} (x, \vec{k}_\perp) \rvert^2 - \lvert B^{\downarrow j, \uparrow}_{uV(ud)} (x, \vec{k}_\perp) \rvert^2 \right), \label{eq:su} \\
  \Delta d(x, \vec{k}_\perp) =&\ \sum_{j=\{-1,0,1\}}
  \left( \lvert B^{\uparrow j,\uparrow}_{dV(uu)} (x, \vec{k}_\perp) \rvert^2 - \lvert B^{\downarrow j,\uparrow}_{dV(uu)} (x, \vec{k}_\perp) \rvert^2 \right), \label{eq:sd}
\end{align}
where $\vec{k}_\perp$ is the transverse momentum. $\Delta u$ and $\Delta d$ can be obtained by integrating $x$ and $\vec{k}_\perp$ away. 
Utilizing the BHL wave function and parameter set II, our model yields $\Delta u = 1.02$, and $\Delta d = -0.24$, giving the spin fraction carried by the quarks as
\begin{equation} \label{eq:spin_fraction}
  \Delta \Sigma = \Delta u + \Delta d = 0.78.
\end{equation}
The value is much larger than the global analysis result $\Delta \Sigma = 0.366$~\cite{deFlorian_global_2008}.
This is not very surprising as our current framework provides only valance contributions.
Sea contribution to magnetic moments is weakened somehow. 
Since Eq.~\eqref{eq:ffc_1} can be interpreted as an integration of charge density for the $Q^2=0$ case, $F_1(0)$ receives no sea quark contribution.
However, no such cancellation happens for the helicity distribution so sea polarization has to be considered in order to give the whole picture.

One possible improvement over this situation is to consider the contribution from the meson clouds. 
In our model, some of the higher-order contributions can be effectively absorbed into the diquark degree of freedom by adjusting the parameters.
Introducing the meson clouds, nevertheless, is still beneficial as there are evidences that the sea quark distribution is flavor-asymmetric~\cite{Ma:2000uv} and polarized~\cite{deFlorian_global_2008}. 
Both the form factors and the spin contents are expected to receive corrections from these effects, which can not be fully included in our wave function as they break the $\mathrm{SU}(6)$ symmetry. 
One similar light-cone model in Ref.~\cite{cloet_nucleon_2012} has attempted such an approach and achieved good agreement with experimental result for the proton spin contents. 

\begin{table}
  \caption{Parameters used in model calculations.\label{tab:par_model}}
  \begin{ruledtabular}
  \begin{tabular}{cddd}
    Quantity  & \multicolumn{1}{c}{Set I (BHL)} & \multicolumn{1}{c}{Set II (BHL)} & \multicolumn{1}{c}{Set III (PL\footnotemark[1])} \\
    \colrule
    $m_u=m_d\,(\GeV)$ & 0.33 &  0.22 & 0.22 \\
    $m_s\,(\GeV )$ & 0.48 & 0.39 & 0.39 \\
    \colrule
    $m_{S(ud)}\,(\GeV)$ & 0.60 & 0.49 & 0.48 \\
    $m_{V(ud)}\,(\GeV)$ & 0.80 & 0.71 & 0.76 \\
    $m_{S(us)}\,(\GeV)$ & 0.75 & 0.85 & 0.80 \\
    $m_{V(us)}\,(\GeV)$ & 0.95 & 1.06 & 1.04 \\
    $m_{V(ss)}\,(\GeV)$ & 1.10 & 1.11 & 1.12 \\
    \colrule
    $\beta_{S(ud)}\,(\GeV)$ & 0.33 & 0.24 & 0.31 \\
    $\beta_{V(ud)}\,(\GeV)$ & 0.33 & 0.31 & 0.32 \\
    $\beta_{S(us)}\,(\GeV)$ & 0.33 & 0.33 & 0.36 \\
    $\beta_{V(us)}\,(\GeV)$ & 0.33 & 0.28 & 0.28 \\
    $\beta_{V(ss)}\,(\GeV)$ & 0.33 & 0.37 & 0.34 \\
  \end{tabular}
  \end{ruledtabular}
  \footnotetext[1]{Power-law wave function.}
\end{table}

\begin{table}
  \caption{Magnetic moments and ratio of axial charges from model calculations. The last two ratios are compared with the BSE calculation in Ref.~\cite{carrillo-serrano_su3-flavour_2014}. Data are from Ref.~\cite{olive_review_2014}.\label{tab:results}}
  \begin{ruledtabular}
    \begin{tabular}{cdddD{+}{\,\pm\,}{-1}}
      Quantity  & \multicolumn{1}{c}{Set I (BHL)} & \multicolumn{1}{c}{Set II (BHL)} & \multicolumn{1}{c}{Set III (PL\footnotemark[1])} & \multicolumn{1}{c}{Expt.\cite{olive_review_2014}/Ref.\cite{carrillo-serrano_su3-flavour_2014}} \\
    \colrule
    $\mu_p$ & 2.21 & 2.82 & 2.79 & 2.792+10^{-8} \\
    $\mu_n$ & -1.26 & -1.70 & -1.69 & -1.913+10^{-7} \\
    $\mu_\Lambda$ & -0.52 & -0.61 & -0.60 & -0.613+0.004 \\
    $\mu_{\Sigma^+}$ & 2.03 & 2.45 & 2.50 & 2.458+0.010 \\
    $\mu_{\Sigma^-}$ & -0.84 & -0.97 & -1.01 & -1.160+0.025 \\
    $|\mu_{\Sigma^0\to\Lambda}|$ & 1.10 & 1.41 & 1.39 & 1.61+0.08 \\
    $\mu_{\Xi^0}$ & -1.10 & -1.26 & -1.28 & -1.250+0.014 \\
    $\mu_{\Xi^-}$ & -0.59 & -0.65 & -0.65 & -0.6507+0.0025 \\
    \colrule
    $g_1/f_1(n\to pe^-\bar{\nu}_e)$ & -1.30 & -1.26 & -1.25 & -1.2723+0.0023 \\
    $g_1/f_1(\Lambda\to pe^-\bar{\nu}_e)$ & -0.82 & -0.83 & -0.81 & -0.718+0.015 \\
    $g_1/f_1(\Sigma^-\to ne^-\bar{\nu}_e)$ & 0.28 & 0.26 & 0.26 & 0.340+0.017 \\
    $f_1/g_1(\Sigma^-\to\Lambda e^-\bar{\nu}_e)$ & <0.01\footnotemark[2] & <0.01\footnotemark[2] & 0.05 & 0.01+0.10 \\
    $g_1/f_1(\Xi^0\to\Sigma^+e^-\bar{\nu}_e)$ & -1.38 & -1.28 & -1.25 & -1.22+0.05 \\
    $g_1/f_1(\Xi^-\to\Lambda e^-\bar{\nu}_e)$ & -0.28 & -0.25 & -0.26 & -0.25+0.05 \\
    $g_1/f_1(\Sigma^-\to\Sigma^0 e^- \bar{\nu}_e)$ & -0.52 & -0.46 & -0.46 & \multicolumn{1}{c}{$-0.44$ \cite{carrillo-serrano_su3-flavour_2014}}\\
    $g_1/f_1(\Xi^-\to\Xi^0 e^- \bar{\nu}_e)$ & 0.27 & 0.23 & 0.23 & \multicolumn{1}{c}{$0.28$ \cite{carrillo-serrano_su3-flavour_2014}} \\
  \end{tabular}
\end{ruledtabular}
\footnotetext[1]{Power-law wave function.}
\footnotetext[2]{The upper limit of absolute values are given here.}
\end{table}

\begin{table}
  \caption{Vector and axial-vector coupling $f_1$ and $g_1$ in semileptonic decays, compared with results in Ref.~\cite{schlumpf_relativistic_1994}. The signs of the form factors in Ref.~\cite{schlumpf_relativistic_1994} are adjusted to meet our choice of baryon wave functions and form factor definition.\label{tab:coupling}}
  \begin{ruledtabular}
    \begin{tabular}{cdddd}
      \null & \multicolumn{2}{c}{Set II (BHL)} & \multicolumn{2}{c}{Ref.~\cite{schlumpf_relativistic_1994}} \\
      \null & \multicolumn{1}{c}{$f_1$} & \multicolumn{1}{c}{$g_1$} & \multicolumn{1}{c}{$f_1$} & \multicolumn{1}{c}{$g_1$}\\
      \colrule
      $n\to pe^-\bar{\nu}_e$ & 1.00 & -1.26 & 1.00 & -1.25 \\
      $\Lambda\to pe^-\bar{\nu}_e$ & 1.18 & -0.98 & 1.04 & -0.79\\
      $\Sigma^-\to ne^-\bar{\nu}_e$ & -0.97 & -0.25 & -0.87 & -0.22\\
      $\Sigma^-\to\Lambda e^-\bar{\nu}_e$ & <0.01\footnotemark & 0.58 & 0.04 & 0.62\\
      $\Xi^0\to\Sigma^+e^-\bar{\nu}_e$ & 0.98 & -1.26 & \multicolumn{1}{c}{--} & \multicolumn{1}{c}{--} \\
      $\Xi^-\to \Lambda e^-\bar{\nu}_e$ & -1.18 & 0.30 & -0.91 & 0.25 \\
      $\Sigma^-\to\Sigma^0 e^- \bar{\nu}_e$ & 1.41 & -0.65 & \multicolumn{1}{c}{--}  & \multicolumn{1}{c}{--} \\
      $\Xi^-\to\Xi^0 e^- \bar{\nu}_e$ & 1.00 & 0.23 & \multicolumn{1}{c}{--}  & \multicolumn{1}{c}{--} \\
    \end{tabular}
  \end{ruledtabular}
  \footnotetext[1]{The upper limit of absolute values are given here.}
\end{table}

\begin{table}
  \caption{Quark sector contributions to the octet magnetic moments. Only valence quark contributions are present in our model.\label{tab:magnetic_moments_flavor}}
  \begin{ruledtabular}
    \begin{tabular}{cdddddd}
      \null & \multicolumn{3}{c}{Set II (BHL)} & \multicolumn{3}{c}{Set III (PL)} \\
      \null & \multicolumn{1}{c}{$\mu^u$} & \multicolumn{1}{c}{$\mu^d$} & \multicolumn{1}{c}{$\mu^s$} & \multicolumn{1}{c}{$\mu^u$} & \multicolumn{1}{c}{$\mu^d$} & \multicolumn{1}{c}{$\mu^s$} \\
      \colrule
      $p$ & 2.64 & 0.19 & \multicolumn{1}{c}{--} & 2.60 & 0.20 & \multicolumn{1}{c}{--} \\
      $n$ & -0.38 & -1.32 & \multicolumn{1}{c}{--} & -0.39 & -1.30 & \multicolumn{1}{c}{--} \\
      $\Lambda$ & 0.09 & -0.04 & -0.65 & 0.14 & -0.07 & -0.64 \\
      $\Sigma^+$ & 2.29 & \multicolumn{1}{c}{--} & 0.17 & 2.34 & \multicolumn{1}{c}{--} & 0.15 \\
      $\Sigma^0$ & 1.14 & -0.57 & 0.17 & 1.17 & -0.58 & 0.15 \\
      $\Sigma^-$ & \multicolumn{1}{c}{--} & -1.14 & 0.17 & \multicolumn{1}{c}{--} & -1.16 & 0.15 \\
      $\Xi^0$ & -0.40 & \multicolumn{1}{c}{--} & -0.86 & -0.41 & \multicolumn{1}{c}{--} & -0.84 \\
      $\Xi^-$ & \multicolumn{1}{c}{--} & 0.20 & -0.86 & \multicolumn{1}{c}{--} & 0.21 & -0.85 \\
    \end{tabular}
  \end{ruledtabular}
\end{table}

\section{Summary\label{sec:summary}}
In summary, we study the electromagnetic and weak form factors of the ground state octet baryons using a light-cone quark-diquark model.
Our model gives a consistent description of the electroweak properties of the baryons in the low momentum transfer region.
Relativistic effects are taken into account by considering the Melosh-Wigner rotation of the quarks and spectator diquarks.
The model results of the baryon magnetic moments and weak charges are in good agreement with experimental as well as with other model calculations.
Two different forms of momentum space wave functions are employed in our calculation.
It is shown that the dependences of magnetic moments and weak charges on the momentum wave functions are small.
We present the Sachs form factors as functions of $Q^2$ up to $4\,\mathrm{GeV}^2$.
Although there is currently no measurement of the form factors except for the nucleon cases, we do notice good agreement between our results and recent lattice results.
Moreover, it is interesting to study other baryon properties within this framework.
Beyond our oversimplified ``quark-diquark'' picture, it is possible to derive the LFWFs from the DSE and light-front holographic~\cite{brodsky_ads/cft_2008} approach, which is also a subject worth exploring.

\begin{acknowledgments}
  The numerical integrations were performed using the CUBA library~\cite{hahn_cuba_2005}. This work is supported by National Natural Science Foundation of China (Grants No.~11035003 and No.~11120101004).
\end{acknowledgments}

\bibliography{octet_ff}

\end{document}